\documentclass[pdflatex,a4paper,12pt]{article}

\usepackage{graphicx}
\usepackage{amsmath,amssymb,amsfonts}
\usepackage{bm}
\usepackage{natbib}
\bibpunct[:]{(}{)}{,}{a}{}{,}
\usepackage[hidelinks]{hyperref}

\begin{document}
\title{Model selection tests for truncated vine copulas under nested hypotheses}

\author{
Ichiro Nishi\textsuperscript{$\dagger$}\thanks{Corresponding author. Email: nishi.ichiro@ism.ac.jp}\ \ and Yoshinori Kawasaki\textsuperscript{$\dagger$} \\ \\\small
\textsuperscript{$\dagger$}The Graduate University for Advanced Studies, The Institute of Statistical Mathematics, \\\small
Midori-cho 10-3, Tachikawa, 190-8562, Tokyo, Japan
}
\date{\empty}
\maketitle
\normalsize
\begin{abstract}
Vine copulas, constructed using bivariate copulas as building blocks, provide a flexible framework for modeling multi-dimensional dependencies. However, this flexibility is accompanied by rapidly increasing complexity as dimensionality grows, necessitating appropriate truncation to manage this challenge. While use of Vuong’s model selection test has been proposed as a method to determine the optimal truncation level, its application to vine copulas has been heuristic, assuming only strictly non-nested hypotheses. This assumption conflicts with the inherent nesting within truncated vine copula structures. In this paper, we systematically apply Vuong's model selection tests to distinguish competing models of truncated vine copulas under both nested and strictly non-nested hypotheses. Through extensive simulation studies, we characterize the conditions under which the nested hypotheses provide improved discernibility and demonstrate that the strictly non-nested framework can still yield valid distinctions in certain settings. This broader perspective on model comparison contributes to both methodological clarity and practical guidance for vine copula truncation.

\end{abstract}

\maketitle

\section{Introduction}

Modeling of statistical dependence through copulas has gained significant attention for its capability to decouple the modeling of univariate marginal distributions based on the dependence structure represented on a uniform scale. In a formal way, a $d$-dimensional copula is defined as a multivariate distribution function on $[0, 1]^d$ with uniform marginal distributions. \cite{Sklar1959} demonstrated that any multivariate distribution can be represented using a copula as follows. Let $\bm X = (X_1, \ldots, X_d)' \sim F$ with marginal distribution functions $F_1, \ldots, F_d$. Then,
\[
F(\bm x) = C(F_1(x_1), \ldots, F_d(x_d)), \quad \bm x := (x_1, \ldots, x_d)',
\]
where $C$ is a $d$-dimensional copula. For a continuous random vector $\bm X$, the copula $C$ is unique. Assuming absolute continuity, the joint density $f$ of $\bm X$ can be expressed as:
\[
f(\bm x) = c(F_1(x_1), \ldots, F_d(x_d)) f_1(x_1) \cdots f_d(x_d),
\]
where $c$ represents the copula density, and $f_1, \ldots, f_d$ are the marginal densities of $f$. Comprehensive discussions on copulas and their properties are provided by \cite{Joe1997} and \cite{Nelsen2006}.

Despite their widespread use, many conventional parametric copula families exhibit limited flexibility, particularly in multi-dimensional contexts. For instance, Archimedean copulas assume exchangeability and are typically parameterized by one or two parameters to define the dependence structure, which may impose restrictive limitations. Elliptical copulas offer enhanced adaptability with their capacity to model varying pairwise dependence, but require specifying large correlation matrices, potentially leading to over-parameterization. Additionally, elliptical copulas inherently possess reflection symmetry, implying identical dependence in both joint lower and upper tails.

A more flexible alternative is provided by vine copulas, a class built on the vine structure introduced by \cite{Joe1996}, with further exploration by \cite{Bedford2001, Bedford2002}. Vine copulas construct multivariate models using a sequence of bivariate copulas distributed over hierarchical levels known as vine trees. This approach allows for the construction of highly flexible multivariate distributions that can effectively handle multi-dimensional dependence structures (see, e.g., \cite{Dissmann2013}, \cite{Brechmann2013}, and \cite{Czado2022}).

However, the enhanced flexibility of vine copulas comes at the cost of increased model complexity, involving a quadratic number of bivariate components, resulting in a potentially large parameter space. Thus, a crucial task is to identify sub-classes of vine copulas that balance data fitting capability with parsimony. One such sub-class is the truncated vine copula, which simplifies the model by considering a limited number of vine trees. Initial investigations into truncated vine copulas were conducted by \cite{Brechmann2012}, who proposed a method for determining the optimal truncation level, i.e., the number of vine trees to include. However, this method is largely heuristic, as it assumes strictly non-nested hypotheses for the competing models while these are nested. An alternative approach proposed by \cite{Brechmann2015} involves using fit indices to select the appropriate truncation level.

The contribution of this paper is to evaluate the performance of Vuong's model selection test under both nested and non-nested hypotheses (Vuong-N and Vuong-SNN tests, respectively) in the context of vine copula truncation. Through extensive simulation studies, we characterize the conditions under which the nested hypotheses provide improved discernibility and demonstrate that the strictly non-nested framework can still yield valid distinctions in certain settings. This broader perspective on model comparison contributes to both methodological clarity and practical guidance for vine copula truncation.

The structure of the paper is as follows. Section 2 introduces and defines vine copulas. Section 3 explores the sub-class of truncated vine copulas. Section 4 reviews Vuong’s model selection tests \citep{Vuong1989} including both Vuong-N and Vuong-SNN test. This section also discusses the truncation level selection methodology based on the Vuong-SNN test, as proposed by \cite{Brechmann2012}. Section 5 provides extensive simulation studies comparing the performance of the Vuong-N and Vuong-SNN tests. Section 6 demonstrates an application that necessitates the Vuong-N test. Finally, Section 7 presents conclusion.

\section{Vine copulas}

Vine copulas utilize various parametric copula families to flexibly model dependencies between pairs of variables. A \(d\)-dimensional vine copula is built from \(\binom{d}{2} = d(d-1)/2\) bivariate copulas. To ensure that this structure forms a valid multivariate copula, specific conditions must be met. Regular vines, introduced by \cite{Bedford2001, Bedford2002}, provide a framework to satisfy these conditions through a sequence of trees. Each tree is a connected acyclic graph, and a regular vine is defined as a sequence of such trees. A set of linked trees \( V = \{T_1, T_2, \ldots, T_{d-1}\} \) constitutes a regular vine (R-vine) on \( d \) elements if:
\begin{enumerate}
    \item \( T_1 \) is a tree with nodes \( N_1 = \{1, \ldots, d\} \) and a set of \( d-1 \) edges denoted by \( E_1 \).
    \item For \( i = 2, \ldots, d-1 \), \( T_i \) is a tree with nodes \( N_i = E_{i-1} \) and edges \( E_i \).
    \item For \( i = 2, \ldots, d-1 \), if \( a = \{a_1, a_2\} \) and \( b = \{b_1, b_2\} \) are nodes in \( N_i \) connected by an edge, exactly one element of \( a \) must match one element of \( b \) (proximity condition).
\end{enumerate}

An R-vine \( V \) consists of \( d(d-1)/2 \) edges across \( d-1 \) trees: \( d-1 \) edges in \( T_1 \), \( d-2 \) in \( T_2 \), down to a single edge in \( T_{d-1} \). Assigning bivariate copulas (pair copulas) to each edge results in an R-vine copula. For formal definitions, we introduce additional notation. The complete union \( A_e \) of an edge \( e = \{a, b\} \in E_i \) in tree \( T_i \) is defined by:
\[
A_e = \{v \in N_1 : \exists e_m \in E_m, m = 1, \ldots, i-1, \text{ such that } v \in e_1 \in \ldots \in e_{i-1} \in e\}.
\]
The conditioning set for edge \( e = \{a, b\} \) is \( D_e := A_a \cap A_b \), and the conditioned sets are \( C_{e,a} := A_a \setminus D_e \) and \( C_{e,b} := A_b \setminus D_e \). \cite{Bedford2001} proved that these conditioned sets are singletons, allowing for edge labels of the form \(\{j(e), k(e) | D(e)\}\). With this, we can define an R-vine copula:

Let \( \bm U = (U_1, \ldots, U_d)' \in [0, 1]^d \) be a random vector with uniform marginal distributions, and \( \bm U_D = \{U_\ell : \ell \in D\} \). Then \(\bm U \) follows the \( d \)-dimensional R-vine copula \( C(\cdot; V, B, \bm\theta) \) if:
\begin{enumerate}
    \item \( V \) is an R-vine on \( d \) elements.
    \item \( B = \{C_{j(e), k(e); D(e)} : e \in E_i, i = 1, \ldots, d-1\} \) is a set of \( d(d-1)/2 \) bivariate copula families representing the conditional distributions of \((U_{j(e)}, U_{k(e)})' | \bm U_{D(e)}\).
    \item \( \bm\theta = \{\bm\theta_{j(e), k(e); D(e)} : e \in E_i, i = 1, \ldots, d-1\} \) are the parameters for these copulas.
\end{enumerate}

A significant advantage of R-vine copulas lies in their tractable density expression, as established by \cite{Bedford2001}. The density of a \( d \)-dimensional R-vine copula \( C(\cdot; V, B, \bm\theta) \) is given by:
\[
c(\bm u; V, B, \bm\theta) = \prod_{i=1}^{d-1} \prod_{e \in E_i} c_{j(e), k(e); D(e)} \left( C_{j(e)|D(e)}(u_{j(e)} | \bm u_{D(e)}), C_{k(e)|D(e)}(u_{k(e)} | \bm u_{D(e)}) \right),
\]
where \( C_{j(e), k(e); D(e)} \) has parameters \( \theta_{j(e), k(e); D(e)} \) and \( C_{\ell|D(e)} \) denotes the conditional distribution of \( U_\ell | \bm U_{D(e)} \), for \( \ell \in \{j(e), k(e)\} \).

Typically, \( C_{j(e), k(e); D(e)} \) depends on the conditioning variables \(\bm U_{D(e)} \) only through its arguments \( C_{j(e)|D(e)}(\cdot | \bm u_{D(e)}) \) and \( C_{k(e)|D(e)}(\cdot | \bm u_{D(e)}) \). Discussions on the simplifying assumption underlying this can be found in \cite{Hobaek2010}, \cite{Acar2012}, and \cite{Stober2013}. The conditional distributions \( C_{\ell|D(e)} \) can be recursively computed tree by tree using the pair copulas from trees \( T_1, \ldots, T_i \) \citep{Dissmann2013}.

\begin{figure}[tbh]
    \begin{center}
        \includegraphics[height=7cm]{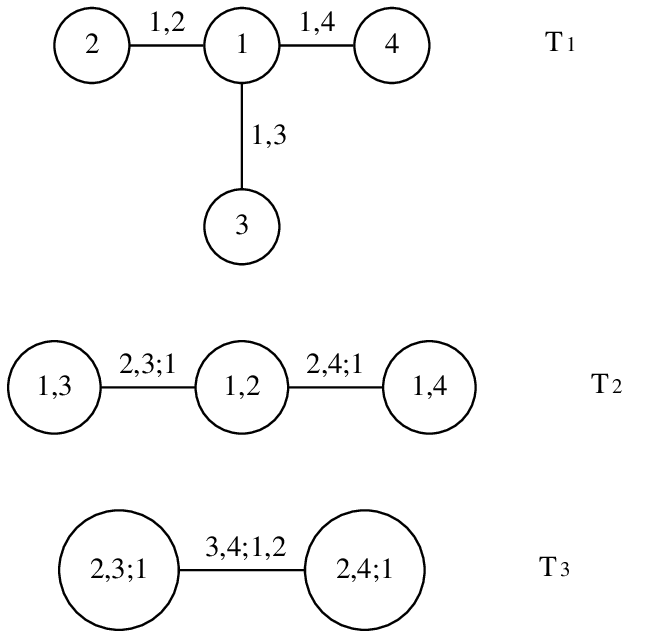}
        \caption{An example of a four-dimensional R-vine \( V \).}
        \label{fig:4d_C}
    \end{center}
\end{figure}

As an example, consider the density of an R-vine copula \( C(\cdot; V, B, \bm\theta) \) corresponding to the R-vine \( V \) illustrated in Fig. \ref{fig:4d_C}. For the pair copulas \( B = B(V) \) and parameters \( \bm\theta = \bm\theta(B(V)) \), the density is:
\[
c(\bm u; V, B, \bm\theta) = c_{1,2}(u_1, u_2) c_{1,3}(u_1, u_3) c_{1,4}(u_1, u_4) c_{2,3;1}(C_{2|1}(u_2 | u_1), C_{3|1}(u_3 | u_1))
\]\vspace*{-1.4cm}

\[
\times c_{2,4;1}(C_{2|1}(u_2 | u_1), C_{4|1}(u_4 | u_1))
c_{3,4;1,2}(C_{3|1,2}(u_3 | u_1, u_2), C_{4|1,2}(u_4 | u_1, u_2))
\]
where \( \bm u = (u_1, \ldots, u_4)' \in [0, 1]^4 \).

The flexibility of R-vine copulas stems from their components: the R-vine structure \( V \), the pair copulas \( B \), and the associated parameters \( \bm\theta \). Different structures and copula choices yield diverse statistical models, forming a rich class of R-vine copulas. As \cite{Morales2011} notes, there are \((d!/2) \times 2^{(d-2)(d-3)/2}\) possible \( d \)-dimensional regular vines, each accommodating \( d(d-1)/2 \) types of copulas and parameters. This vast space of options motivates the study of more tractable sub-classes, such as truncated R-vines.

\section{Truncation}

Truncated R-vine copulas were initially introduced by \cite{Brechmann2012} to simplify multi-dimensional dependency structures by truncating the vine structure at a certain tree level. Formally, consider a random vector \(\bm U = (U_1, \ldots, U_d)'\) with uniform marginal distributions, and let \( \ell \in \{0, \ldots, d - 1\} \) represent the truncation level. The vector \(\bm U\) is said to follow a \(d\)-dimensional \(\ell\)-truncated R-vine copula, denoted as \(tRV(\ell)\), if it is a \(d\)-dimensional R-vine copula where:
\[
C_{j(e), k(e); D(e)} = \Pi \quad \forall e \in E_i, \, i = \ell + 1, \ldots, d - 1,
\]
where \(\Pi\) denotes the bivariate independence copula.

Truncation significantly reduces the complexity of the model because the dependence structure beyond tree \(T_\ell\) is assumed to be independent, leading to a reduced number of pair copulas:
\[
\sum_{i=1}^{\ell} (d - i) = \frac{\ell(2d - (\ell + 1))}{2},
\]
which scales linearly with \(d\) for a fixed \(\ell\). Notable special cases include \(\ell = 0\), which corresponds to a multivariate independence copula; \(\ell = 1\), representing a Markov tree model; and \(\ell = d - 1\), which results in a fully specified R-vine copula. As the density of the independence copula is unity, the density of a truncated R-vine copula simplifies to:
\[
c(\bm u; V, B, \bm\theta) = \prod_{i=1}^{\ell} \prod_{e \in E_i} c_{j(e), k(e); D(e)} \left( C_{j(e)|D(e)}(u_{j(e)} | \bm u_{D(e)}), C_{k(e)|D(e)}(u_{k(e)} | \bm u_{D(e)}) \right).
\]
This simplification facilitates parameter estimation using likelihood-based methods.

Justification for using truncated R-vine copulas is discussed in \cite{Brechmann2012, Dissmann2013}. They argue that the essential dependencies among variables are typically captured by the pair copulas in the initial trees. Furthermore, \cite{Joe2010} highlight that if the pair copulas in the first tree possess tail dependence, the overall model can exhibit tail dependence. Additionally, truncation helps mitigate rounding errors in recursive calculations of conditional distributions used as arguments in pair copula densities. Thus, truncated R-vine copulas maintain a balance between parsimony and the flexibility of full R-vine copulas.

Choosing an appropriate truncation level \(\ell\) involves finding a balance where the \(\ell\)-truncated R-vine copula adequately fits the data while maintaining parsimony. The goal is to select the optimal model from all possible \(\ell\)-truncated R-vine copulas that balances simplicity and flexibility.

\section{Vuong test}

In this section, we first briefly review the theory of the Vuong test \citep{Vuong1989}. We consider two parametric families of conditional distributions for \( Y_t \) given \( Z_t \):
\[
\bm F_{\bm\theta} := \{ F_{Y | Z}(\cdot | \cdot; \bm\theta) : \bm\theta \in \bm\Theta \subset \mathbb{R}^p \} \quad \text{and} \quad \bm G_{\bm \gamma} := \{ G_{Y | Z}(\cdot | \cdot; \bm\gamma) : \bm\gamma \in \bm\Gamma \subset \mathbb{R}^q \}.
\]
No specific relationship is initially imposed between the models \( \bm F_{\bm\theta} \) and \( \bm G_{\bm\gamma} \), allowing for cases where they are nested, overlapping, or strictly non-nested. One, both, or neither model may correctly describe the true conditional distribution of \( Y_t \) given \( Z_t \).

Assuming standard regularity conditions, we define the following matrices for \( \bm F_{\bm\theta} \):
\[
A_f(\bm\theta) := E^0 \left[ \frac{\partial^2 \log f(Y_t | Z_t;\bm \theta)}{\partial \bm\theta \partial \bm\theta'} \right],
\]
\[
B_f(\bm\theta) := E^0 \left[ \frac{\partial \log f(Y_t | Z_t; \bm\theta)}{\partial \bm\theta} \cdot \frac{\partial \log f(Y_t | Z_t; \bm\theta)}{\partial \bm\theta'} \right],
\]
where \(E^0[\cdot]\) denotes the expectation with respect to the true distribution of \(X_t = (Y_t, Z_t)\). Analogous matrices \(A_g(\bm\gamma)\) and \(B_g(\bm\gamma)\) are defined for \(\bm G_{\bm\gamma}\), as well as:
\[
B_{fg}(\bm\theta, \bm\gamma) = B_{gf}'(\bm\gamma, \bm\theta) := E^0 \left[ \frac{\partial \log f(Y_t | Z_t; \bm\theta)}{\partial \bm\theta} \cdot \frac{\partial \log g(Y_t | Z_t; \bm\gamma)}{\partial \bm\gamma} \right].
\]

Let \(\bm\theta_*\) and \(\bm\gamma_*\) denote the pseudo-true values of \(\bm\theta\) and \(\bm\gamma\) for \(\bm F_{\bm\theta}\) and \(\bm G_{\bm\gamma}\), respectively (see, e.g., \cite{Sawa1978}). The likelihood ratio (LR) statistic for \(\bm F_{\bm\theta}\) versus \(\bm G_{\bm\gamma}\) is given by:
\[
LR_n\left(\hat{\bm\theta}_n,\hat{\bm\gamma}_n\right) := \sum_{t=1}^n \log \frac{f(Y_t | Z_t; \hat{\bm\theta}_n)}{g(Y_t | Z_t; \hat{\bm\gamma}_n)},
\]
where \(\hat{\bm\theta}_n\) and \(\hat{\bm\gamma}_n\) are the maximum likelihood estimators of \(\bm\theta_*\) and \(\bm\gamma_*\).

Let \(Z = (Z_1, \ldots, Z_m)'\) be a vector of \(m\) independent standard normal variables, and let \(\lambda = (\lambda_1, \ldots, \lambda_m)'\) be a vector of \(m\) real numbers. Then, \(\sum_{i=1}^m \lambda_i Z_i^2\) is distributed as a weighted sum of chi-square variables, with the cumulative distribution function denoted by \(M_m(\cdot; \lambda)\).

If \(f(\cdot|\cdot; \bm\theta_*) = g(\cdot|\cdot; \bm\gamma_*)\), then
\[
2LR_n\left( \hat{\bm\theta}_n,\hat{\bm\gamma}_n \right) \xrightarrow{\;\;D\;\;} M_{p+q}(\cdot;\lambda_*),
\]
where \(\lambda_*\) is the vector of \(p + q\) (potentially negative) eigenvalues of
\[
W=\begin{bmatrix}
-B_f(\bm\theta_*) A_f^{-1}(\bm\theta_*) & -B_{fg}(\bm\theta_*, \bm\gamma_*) A_g^{-1}(\bm\gamma_*) \\
B_{gf}(\bm\gamma_*,\bm\theta_*) A_f^{-1}(\bm\theta_*) & B_g(\bm\gamma_*) A_g^{-1}(\bm\gamma_*)
\end{bmatrix}.
\]
The condition \(f(\cdot|\cdot; \bm\theta_*) = g(\cdot|\cdot; \bm\gamma_*)\) indicates that the models \(\bm F_{\bm\theta}\) and \(\bm G_{\bm\gamma}\) are observationally equivalent in their closest representation of the true distribution.

If \(f(\cdot|\cdot; \bm\theta_*) \neq g(\cdot|\cdot; \bm\gamma_*)\), then
\[
n^{-1/2}LR_n\left( \hat{\bm\theta}_n,\hat{\bm\gamma}_n \right)-n^{1/2}E^0\left[\log \frac{f(Y_t | Z_t; \bm\theta_*)}{g(Y_t | Z_t; \bm\gamma_*)}\right] \xrightarrow{\;\;D\;\;} N(0, \omega_*^2),
\]
where \(\omega_*\) is the variance of \(\log \frac{f(Y_t | Z_t; \bm\theta_*)}{g(Y_t | Z_t; \bm\gamma_*)}\), where the variance is calculated with respect to the true distribution of \(X_t = (Y_t, Z_t)\). That is:
 \[
\omega_*^2:=Var^0\left[\log \frac{f(Y_t | Z_t; \bm\theta_*)}{g(Y_t | Z_t; \bm\gamma_*)}\right]
=E^0\left[\log \frac{f(Y_t | Z_t; \bm\theta_*)}{g(Y_t | Z_t; \bm\gamma_*)}\right]^2
-\left[E^0\left[\log \frac{f(Y_t | Z_t; \bm\theta_*)}{g(Y_t | Z_t; \bm\gamma_*)}\right]\right]^2
\]

We consider the following hypotheses and definitions:
\[
H_0: E^0\left[ \log \frac{f(Y_t | Z_t; \bm\theta_*)}{g(Y_t | Z_t; \bm\gamma_*)} \right] = 0,
\]
which states that \(\bm F_{\bm\theta}\) and \(\bm G_{\bm\gamma}\) are equivalent, against
\[
H_f: E^0\left[ \log \frac{f(Y_t | Z_t; \bm\theta_*)}{g(Y_t | Z_t; \bm\gamma_*)} \right] > 0,
\]
which states that \(\bm F_{\bm\theta}\) is superior to \(\bm G_{\bm\gamma}\), or
\[
H_g: E^0\left[ \log \frac{f(Y_t | Z_t; \bm\theta_*)}{g(Y_t | Z_t; \bm\gamma_*)} \right] < 0,
\]
which states that \(\bm F_{\bm\theta}\) is inferior to \(\bm G_{\bm\gamma}\).

We now consider the case where \(\bm F_{\bm\theta} \cap \bm G_{\bm\gamma} = \phi\), i.e., for strictly non-nested models. Since the models \(\bm F_{\bm\theta}\) and \(\bm G_{\bm\gamma}\) do not have any distribution in common, \(f(\cdot|\cdot; \bm\theta_*) \neq g(\cdot|\cdot; \bm\gamma_*)\) is fulfilled. Under $H_0$, the asymptotic variance $\omega_*^2$ can be consistently estimated by:
\[
\hat{\omega}_n^2:=\frac{1}{n}\sum_{t=1}^n \left[\log \frac{f(Y_t | Z_t; \hat{\bm\theta}_n)}{g(Y_t | Z_t; \hat{\bm\gamma}_n)}\right]^2 - \left[\frac{1}{n}\sum_{t=1}^n \log \frac{f(Y_t | Z_t; \hat{\bm\theta}_n)}{g(Y_t | Z_t; \hat{\bm\gamma}_n)}\right]^2 .
\]
We obtain the following likelihood ratio test for strictly non-nested models:
\begin{enumerate}
    \item Under \(H_0\), \(n^{-1/2}LR_n\left( \hat{\bm\theta}_n,\hat{\bm\gamma}_n \right)/\hat\omega_n \xrightarrow{\;\;D\;\;} N(0,1),\)
    \item under \(H_f\), \(n^{-1/2}LR_n\left( \hat{\bm\theta}_n,\hat{\bm\gamma}_n \right)/\hat\omega_n \xrightarrow{\;a.s.\;} +\infty,\)
    \item under \(H_g\), \(n^{-1/2}LR_n\left( \hat{\bm\theta}_n,\hat{\bm\gamma}_n \right)/\hat\omega_n \xrightarrow{\;a.s.\;} -\infty.\)
\end{enumerate}
We choose a critical value $c$ from the standard normal distribution for some significance level. If the value of the statistic $n^{-1/2}LR_n( \hat{\bm\theta}_n,\hat{\bm\gamma}_n )/\hat\omega_n$ is larger than $c$ then we reject $H_0$ in favor of \(\bm F_{\bm\theta}\) being superior to \(\bm G_{\bm\gamma}\). If  $n^{-1/2}LR_n( \hat{\bm\theta}_n,\hat{\bm\gamma}_n )/\hat\omega_n$ is smaller than $-c$ then we reject $H_0$ in favor of \(\bm G_{\bm\gamma}\) being superior to \(\bm F_{\bm\theta}\). Finally if $|n^{-1/2}LR_n( \hat{\bm\theta}_n,\hat{\bm\gamma}_n )/\hat\omega_n|\leq c$ then we cannot discriminate between the two competing models given the data. The test is applicable regardless of whether both, one, or neither model is misspecified. We call this type of Vuong's model selection test as {\bf Vuong-SNN test}, hereafter, following the original expression in \cite{Vuong1989} of ``Tests for {\bf S}trictly {\bf N}on-{\bf N}ested Models''.

Next, we consider the case where \(\bm G_{\bm\gamma} \subset \bm F_{\bm\theta}\), i.e., for nested models. The alternative to the null hypothesis $H_0$, denoted as $H_A$, is $H_f$ since $H_g$ can never occur because $\bm G_{\bm\gamma}$ can never superior to $\bm F_{\bm\theta}$. Based on the result for the condition \(f(\cdot|\cdot; \bm\theta_*) = g(\cdot|\cdot; \bm\gamma_*)\), we obtain the following likelihood ratio test for nested models:
\begin{enumerate}
    \item Under \(H_0\), \(2LR_n\left( \hat{\bm\theta}_n,\hat{\bm\gamma}_n \right) \xrightarrow{\;\;D\;\;} M_{p+q}(\cdot;\lambda_*),\)
    \item under \(H_A\), \(2LR_n\left( \hat{\bm\theta}_n,\hat{\bm\gamma}_n \right) \xrightarrow{\;a.s.\;} +\infty.\)
\end{enumerate}
The test is one-sided and is conducted by choosing a critical value from \(M_{p+q}(\cdot;\lambda_*)\) and rejecting the null hypothesis that the models are equivalent if \(2LR_n( \hat{\bm\theta}_n,\hat{\bm\gamma}_n )\) exceeds this critical value. The test is applicable regardless of whether the larger model is correctly specified. We call this type of Vuong's model selection test as {\bf Vuong-N test}, hereafter, following the original expression in \cite{Vuong1989} of ``Tests for {\bf N}ested Models''.

\cite{Brechmann2012} suggest a sequential approach for determining the truncation level. They build models incrementally, starting from a 1-truncated copula and proceeding to higher-order truncated copulas, using the Vuong-SNN test at each stage to assess the benefit of including an additional tree. If the additional complexity does not yield significant improvement, the truncation is finalized at the current level.
The test statistic is based on the standardized sum \(\nu\) of log-likelihood differences \(m_i := \log \left[ \frac{f(x_i | \hat{\bm\theta})}{g(x_i | \hat{\bm\gamma})} \right]\) for observations \(x_i\), \(i = 1, \ldots, n\). Thus, model \(f\) is preferred over \(g\) at significance level \(\alpha\) if:
\[
\nu := \frac{1}{n} \sum_{i=1}^n m_i \bigg/ \sqrt{\frac{1}{n} \sum_{i=1}^n (m_i - \bar{m})^2} > z_{1 - \alpha/2},
\]
where \(z_{1 - \alpha/2}\) is the \((1 - \alpha/2)\) quantile of the standard normal distribution. If \(\nu < -z_{1 - \alpha/2}\), model \(g\) is chosen. If \(|\nu| \leq z_{1 - \alpha/2}\), no distinction can be made between the models.

\cite{Brechmann2012} heuristically apply the Vuong-SNN test to compare \(tRV(\ell)\) and \(tRV(\ell + 1)\), even though these models are nested. They conclude that the Vuong-SNN test performs effectively in selecting the appropriate truncation level for R-vine copulas. However, the performance of model distinction by Vuong-N test in vine copula truncation settings has not been reported and this is the contribution of this paper.

\section{Simulation studies}

\begin{figure}[tbh]
    \begin{center}
        \vspace*{0.5cm}
        \includegraphics[height=3cm]{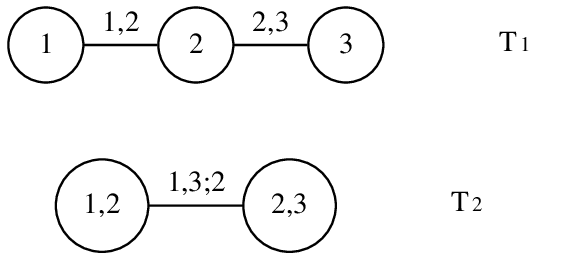}
        \caption{Three-dimensional R-vine structure used in the simulation studies.}
        \label{fig:3d}
    \end{center}
\end{figure}

We conducted both the Vuong-N and Vuong-SNN tests to compare their performance of distinguishing competing models of truncated vine copulas. The initial analysis focused on a three-dimensional scenario, illustrated in Fig. \ref{fig:3d}.

\begin{itemize}
    \item \textbf{True Models \((h_0)\):} Gaussian copulas were used in trees \(T_1\) and \(T_2\) with Kendall’s \(\tau\) values \(\tau_{T_1}\) and \(\tau_{T_2}\) varying across different levels of dependence. Specifically, the \(\tau_{T_1}\) and \(\tau_{T_2}\) values ranged from \(0.04\) to \(0.28\) in increments of \(0.04\), resulting in 49 distinct patterns of true models.
    \item \textbf{Larger Models \((\bm F_{\bm\theta})\):} Each copula \(c_{1,2}\), \(c_{2,3}\), and \(c_{1,3;2}\) was specified as Gaussian.
    \item \textbf{Truncated Models \((\bm G_{\bm\gamma})\):} The copulas \(c_{1,2}\) and \(c_{2,3}\) were Gaussian, while the copula \(c_{1,3;2}\) was represented by an independence copula.
\end{itemize}

\begin{figure}[tbh]
    \begin{center}
        \includegraphics[height=6cm]{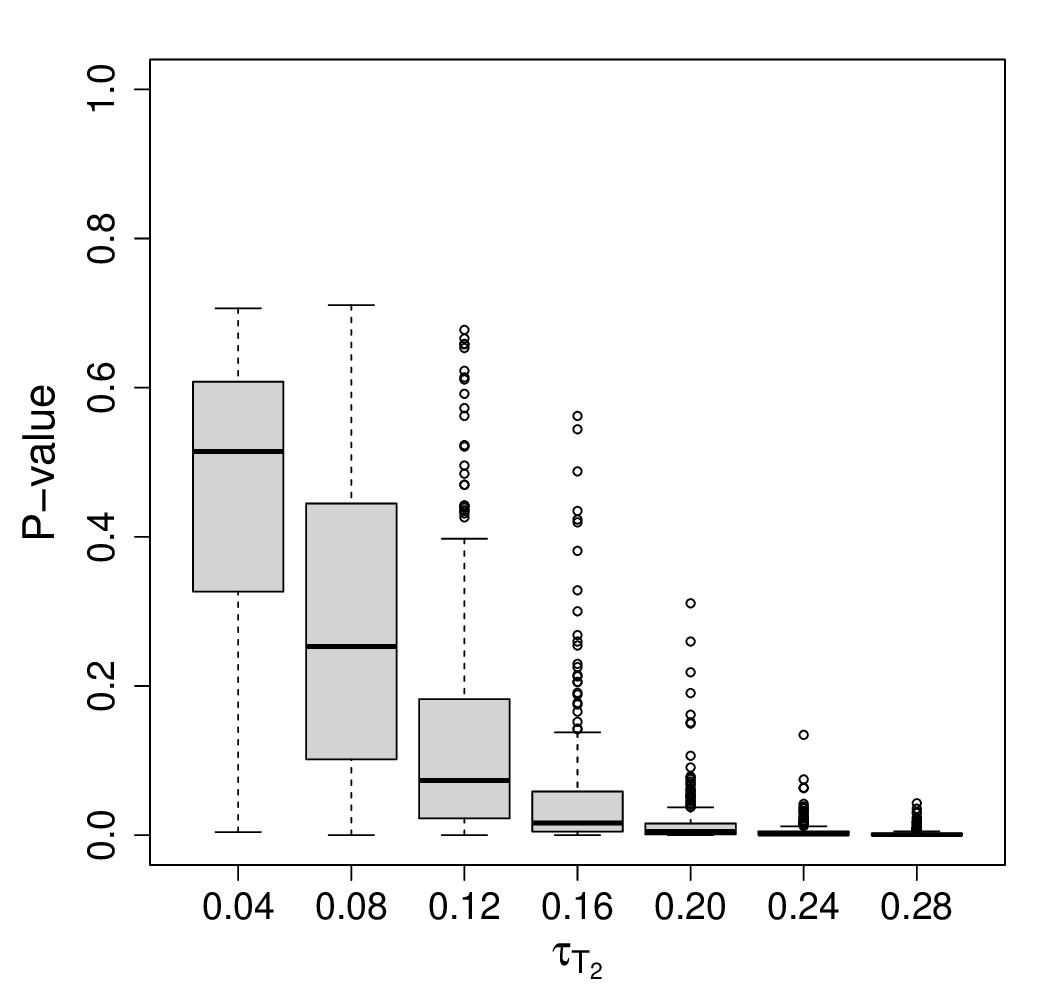}
        \includegraphics[height=6cm]{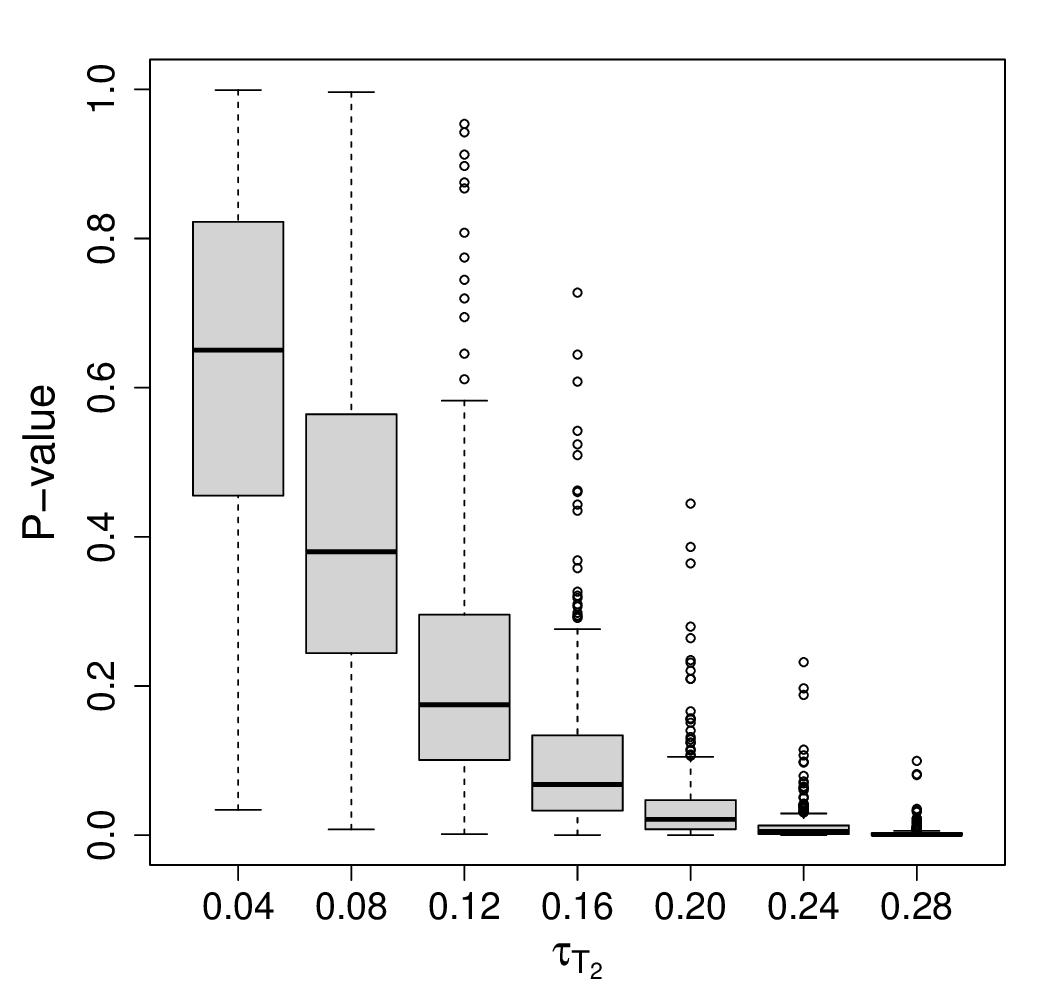}\\
        \includegraphics[height=6cm]{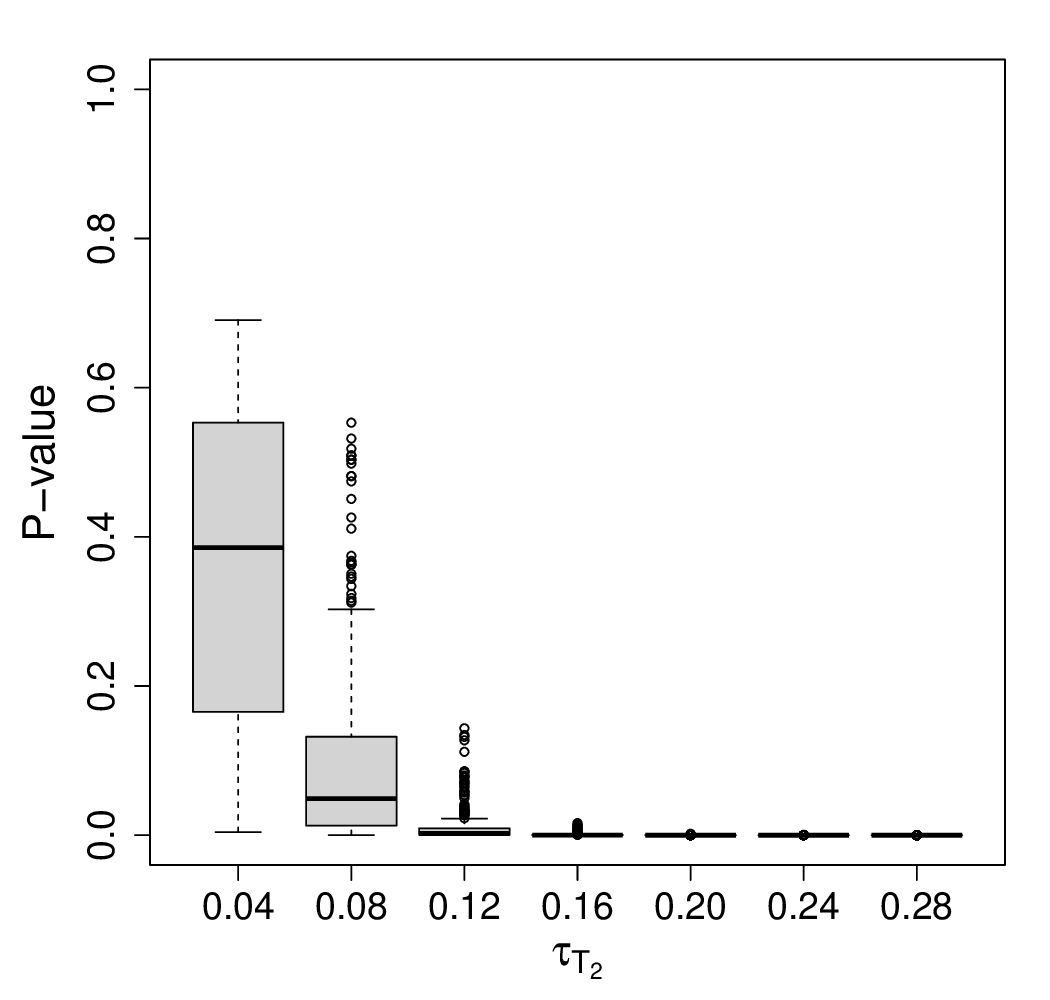}
        \includegraphics[height=6cm]{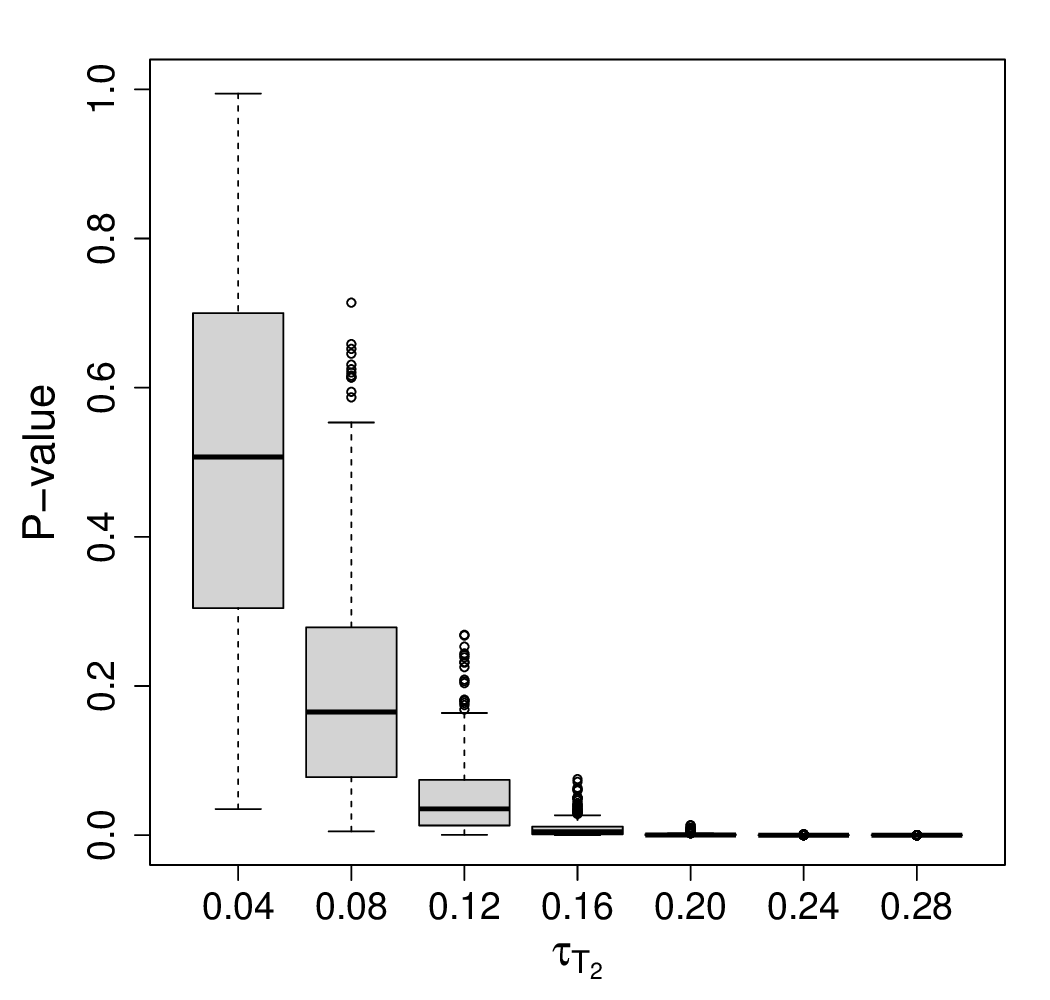}
        \caption{P-values obtained by Vuong-N (left) and Vuong-SNN (right) tests comparing $\bm F_{\bm\theta}$ and $\bm G_{\bm\gamma}$ for \(\tau_{T_1} = 0.20\) with \(n = 200\) (top) and \(n = 500\) (bottom). The corresponding \(\tau_{T_2}\) values are as shown.}
        \label{fig:3d_pval}
    \end{center}
\end{figure}

For each setting, we simulated \(n \in \{100, 200, 500, 1000\}\) observations from the true models. We then optimized the models \(\bm{F}_{\bm{\theta}}\) and \(\bm{G}_{\bm{\gamma}}\) to the simulated data, using the true parameters as starting values, to obtain the pseudo-true parameters \(\bm{\theta}_*\) and \(\bm{\gamma}_*\) via the R package \texttt{VineCopula} \citep{VineCopula}. P-values were computed by Vuong-N and Vuong-SNN tests using the R package \texttt{nonnest2} \citep{nonnest2}. For calculating empirical $A_f(\bm \theta_*)$ and $A_g(\bm \gamma_*)$ we apply the methodology of quasi-maximum likelihood estimation proposed by \cite{White1982} (see, for example, \citet[Section 5.8]{Hamilton1994}). Each scenario was repeated \(R = 300\) times, and box plots of the p-values are provided for \(\tau_{T_1} = 0.20\) and \(n = 200, 500\) as examples in Fig. \ref{fig:3d_pval}.

In these results, smaller p-values are favorable as they indicate models that better fit the data, given \(\tau_{T_2} \neq 0\) in all true model patterns. With larger \(n\), p-values tended to shift closer to 0, reflecting increased power to distinguish between the two models. Comparing Vuong-N and Vuong-SNN tests, p-values obtained by Vuong-N test were consistently lower. This difference is particularly notable for \(n = 500\) and \(\tau_{T_2} = 0.08\), where the median p-value obtained by Vuong-N test was \(0.049\) compared to \(0.17\) obtained by Vuong-SNN test.

\begin{figure}[tbh]
    \begin{center}
        \includegraphics[height=6cm]{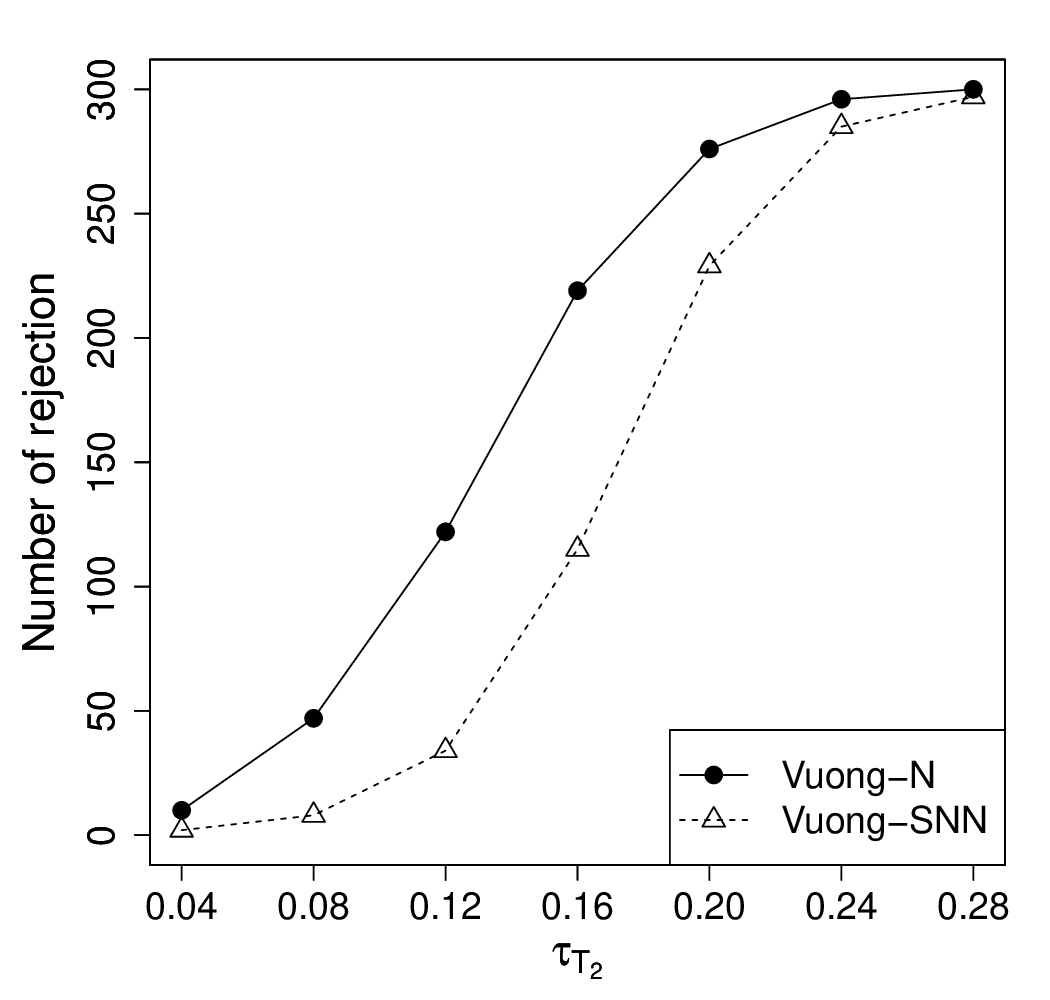}
        \includegraphics[height=6cm]{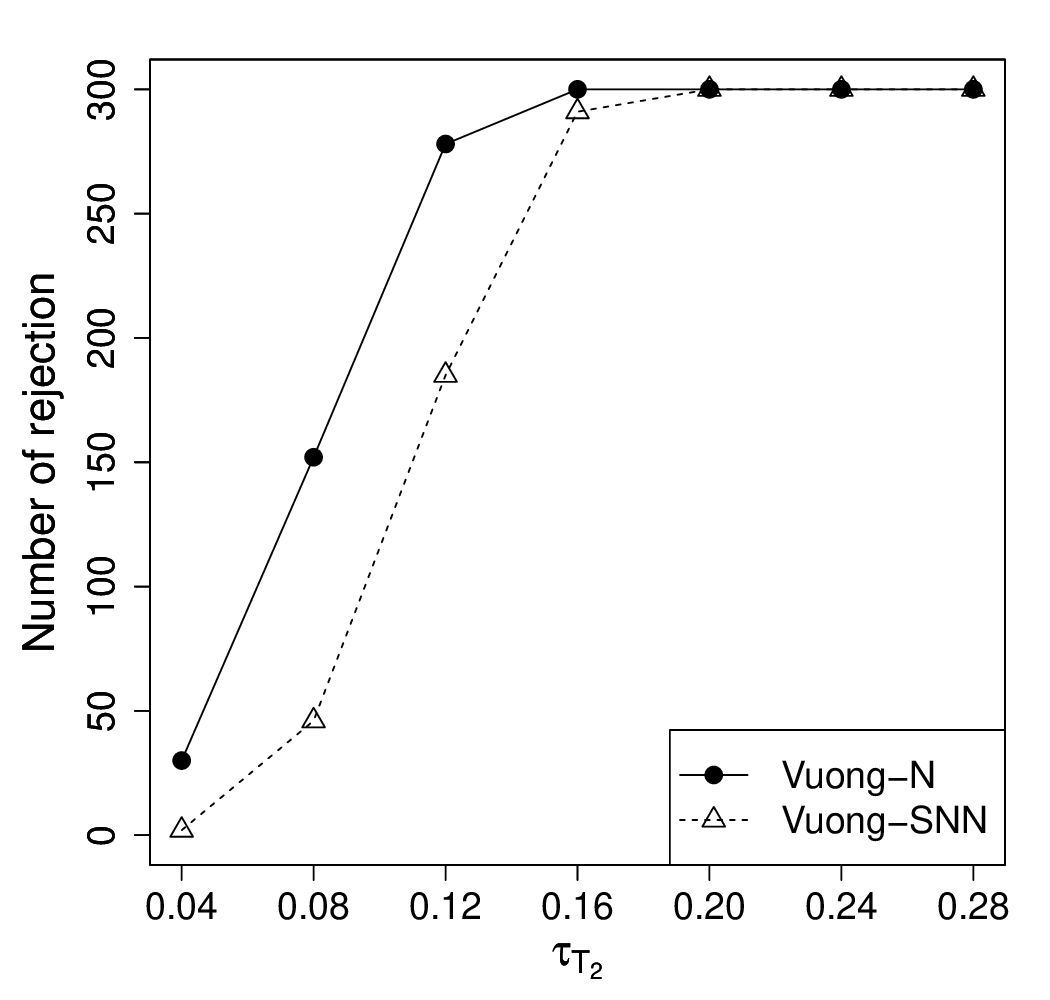}\\
        \includegraphics[height=6cm]{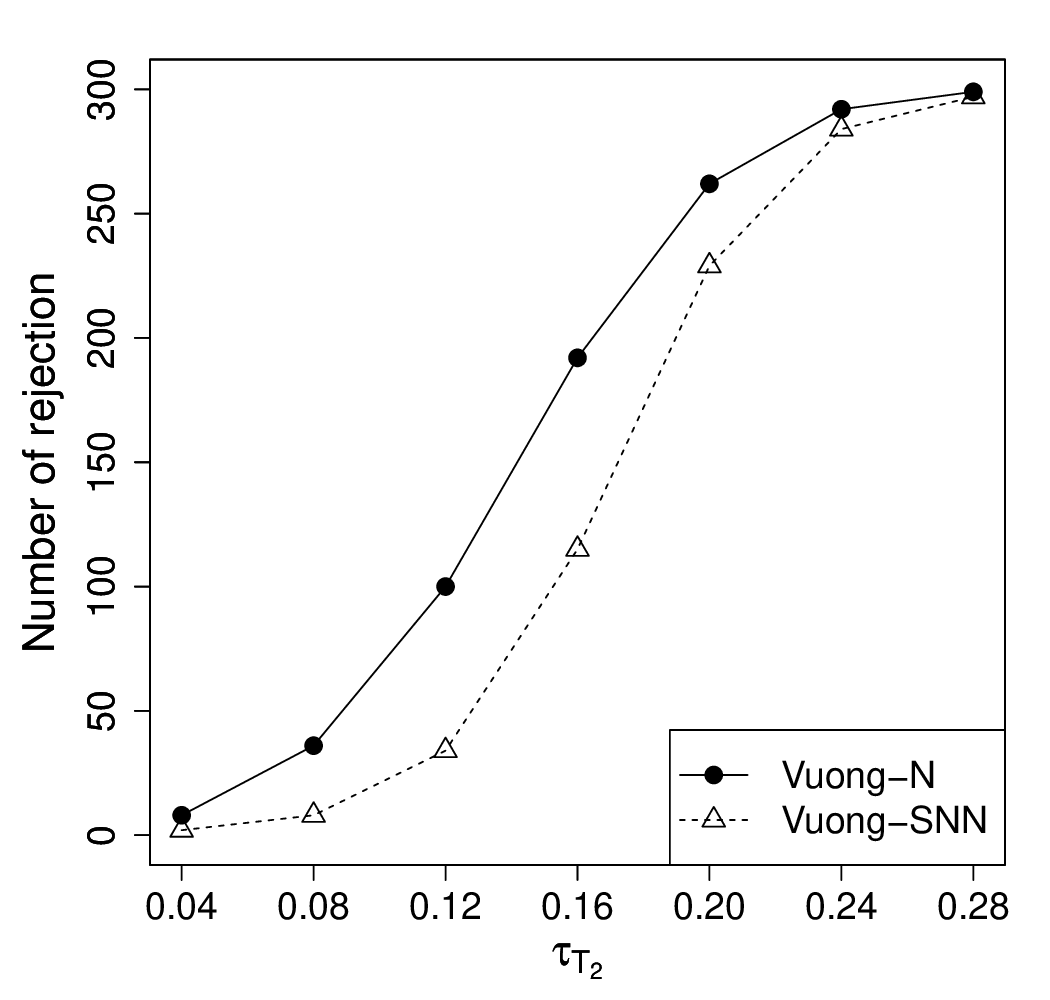}
        \includegraphics[height=6cm]{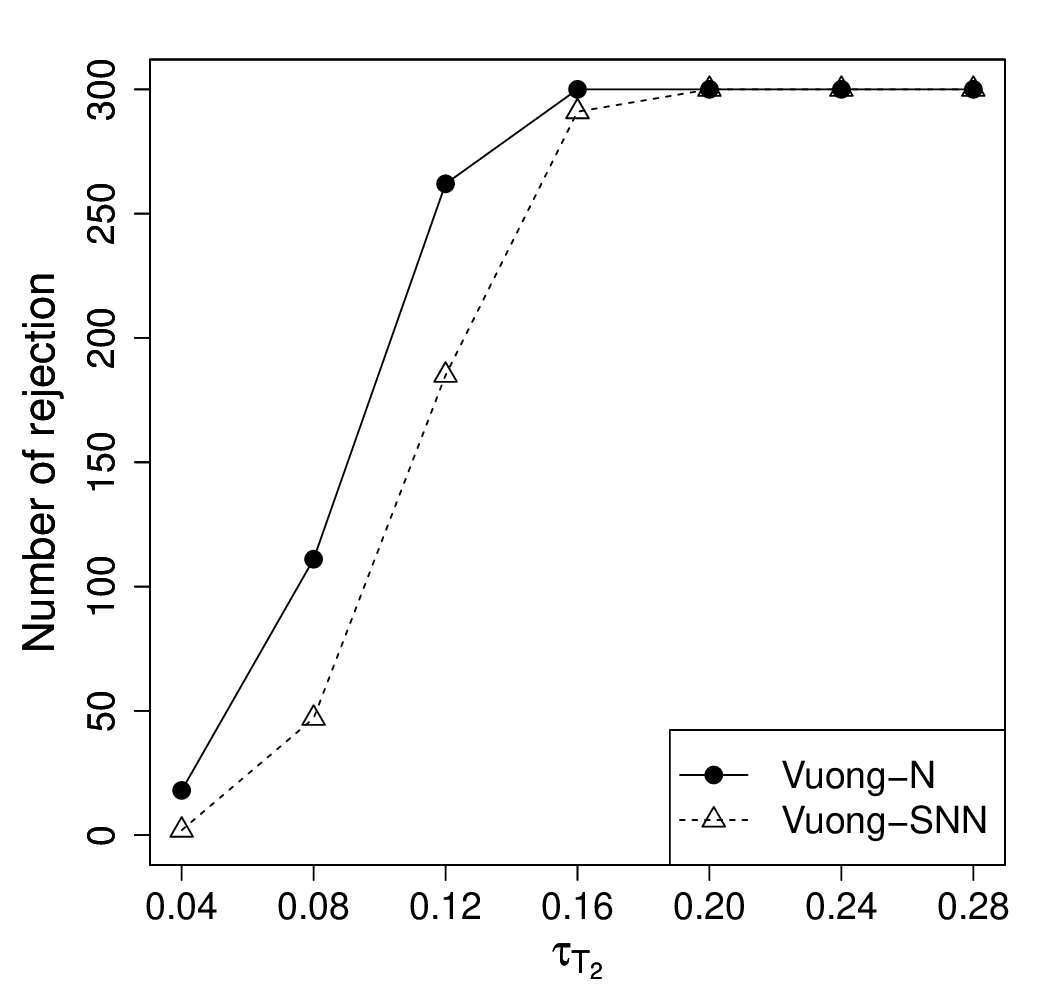}
        \caption{The number of rejections by the Vuong test $(\alpha=0.05)$ comparing \(\bm F_{\bm\theta}\) and \(\bm G_{\bm\gamma}\) for \(\tau_{T_1} = 0.20\) (top), corresponding to the conditions in Fig. \ref{fig:3d_pval}, and \(\tau_{T_1} = 0.28\) (bottom) with \(n = 200\) (left) and \(n = 500\) (right).} 
        \label{fig:3d_num}
    \end{center}
\end{figure}

Using a significance level of \(\alpha = 0.05\), we counted the rejections. The results for the settings in Fig. \ref{fig:3d_pval} as well as the case where $\tau_{T_1} = 0.28$ are shown in Fig. \ref{fig:3d_num}. Vuong-N test yielded more rejections of \(H_0\) than Vuong-SNN test, especially in cases with weaker dependencies around \(\tau_{T_2} \approx 0.12\). The differences diminished as \(\tau_{T_2}\) increased. For \(\tau_{T_2} \approx 0.04\), both tests struggled to differentiate the two models. Similarly, as \(\tau_{T_1}\) increased, the difference between rejection rates decreased.

\begin{figure}[tbh]
    \begin{center}
        \includegraphics[height=6cm]{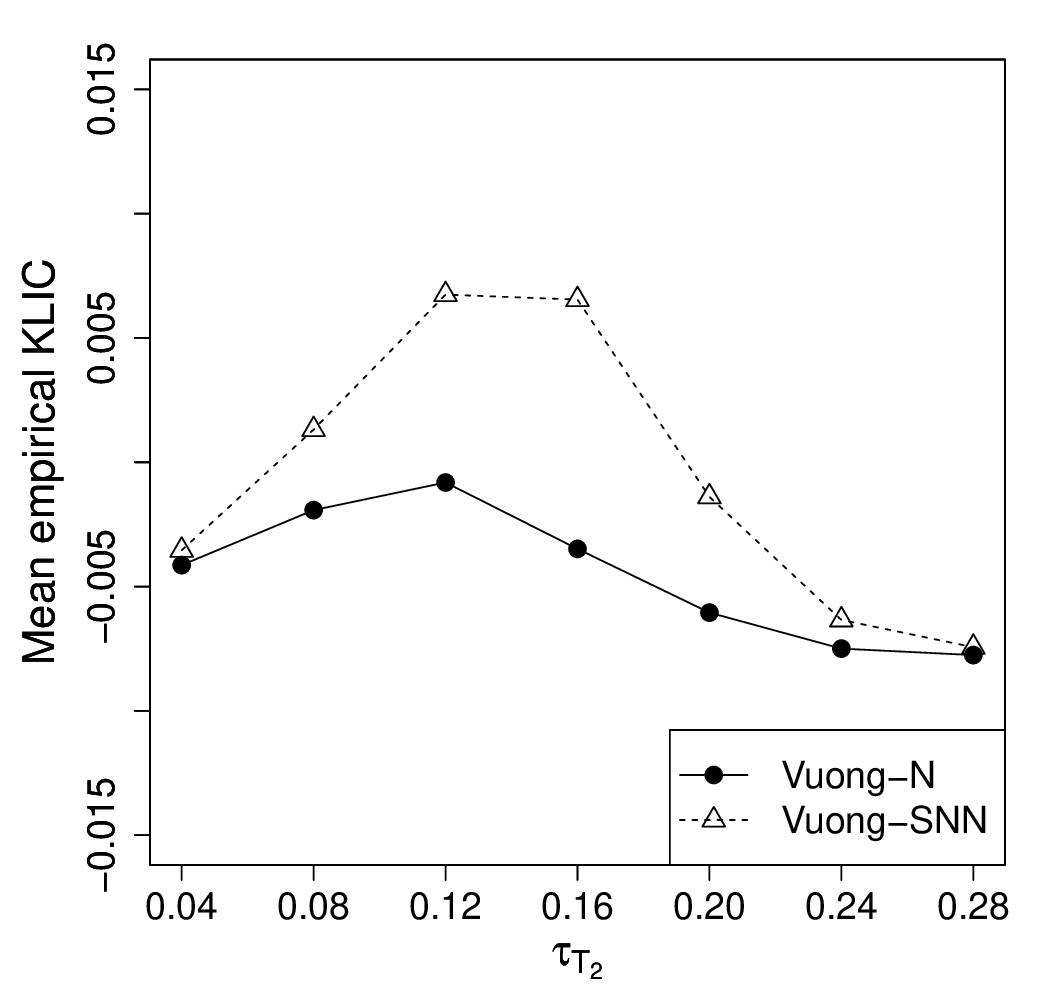}
        \includegraphics[height=6cm]{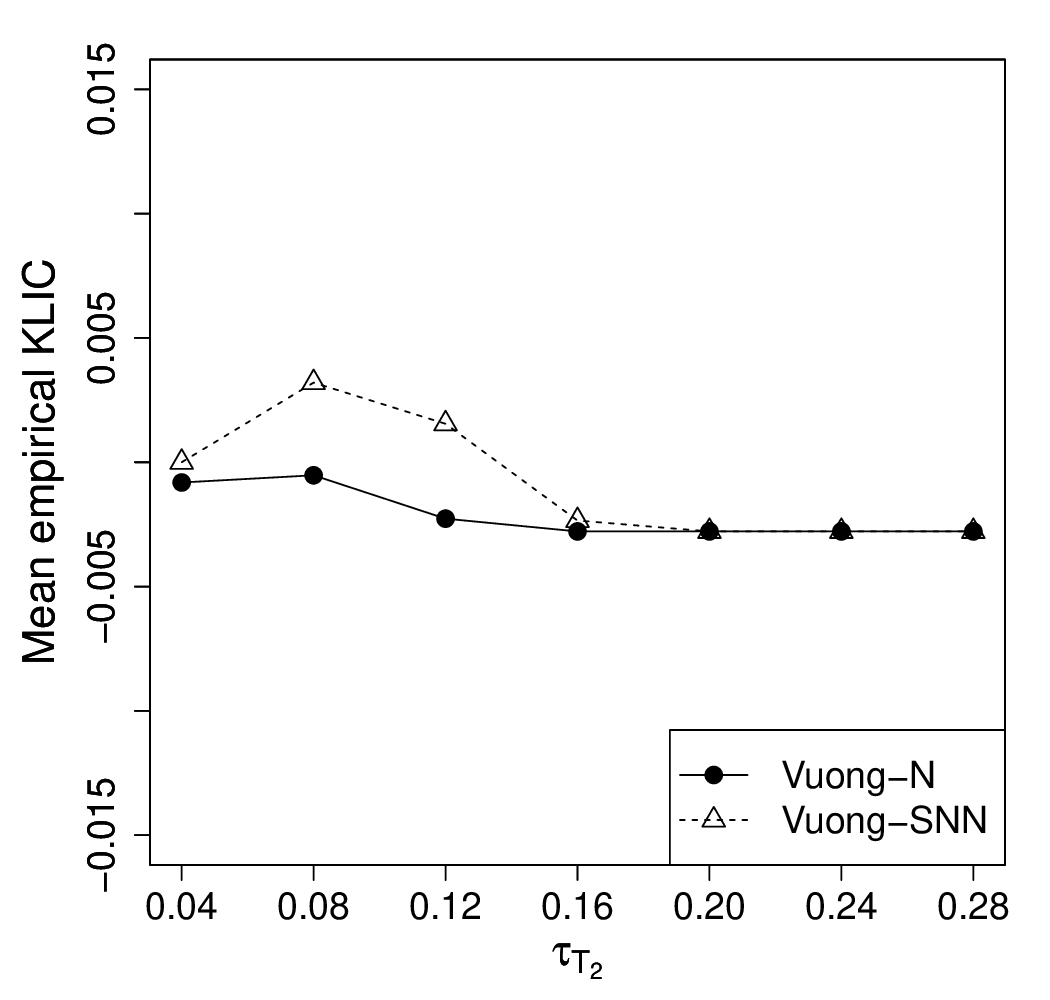}\\
        \includegraphics[height=6cm]{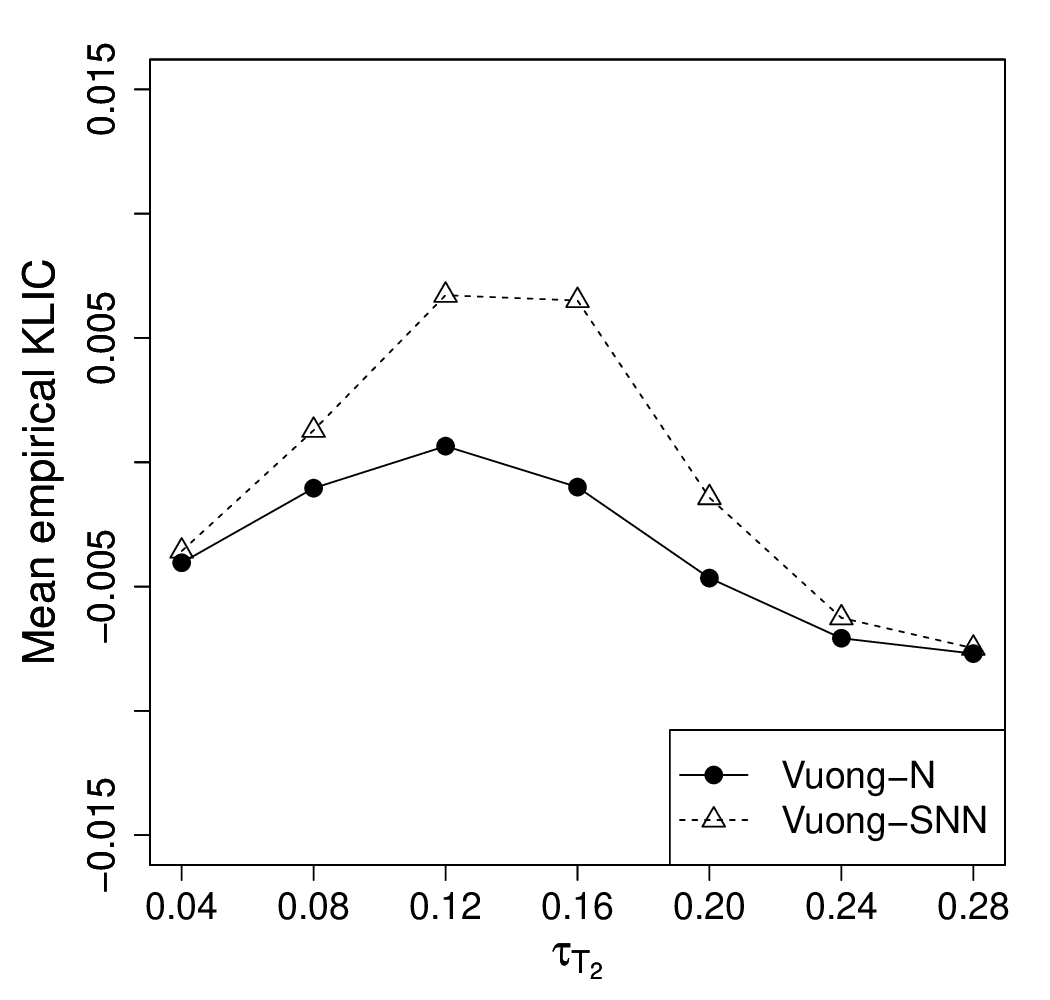}
        \includegraphics[height=6cm]{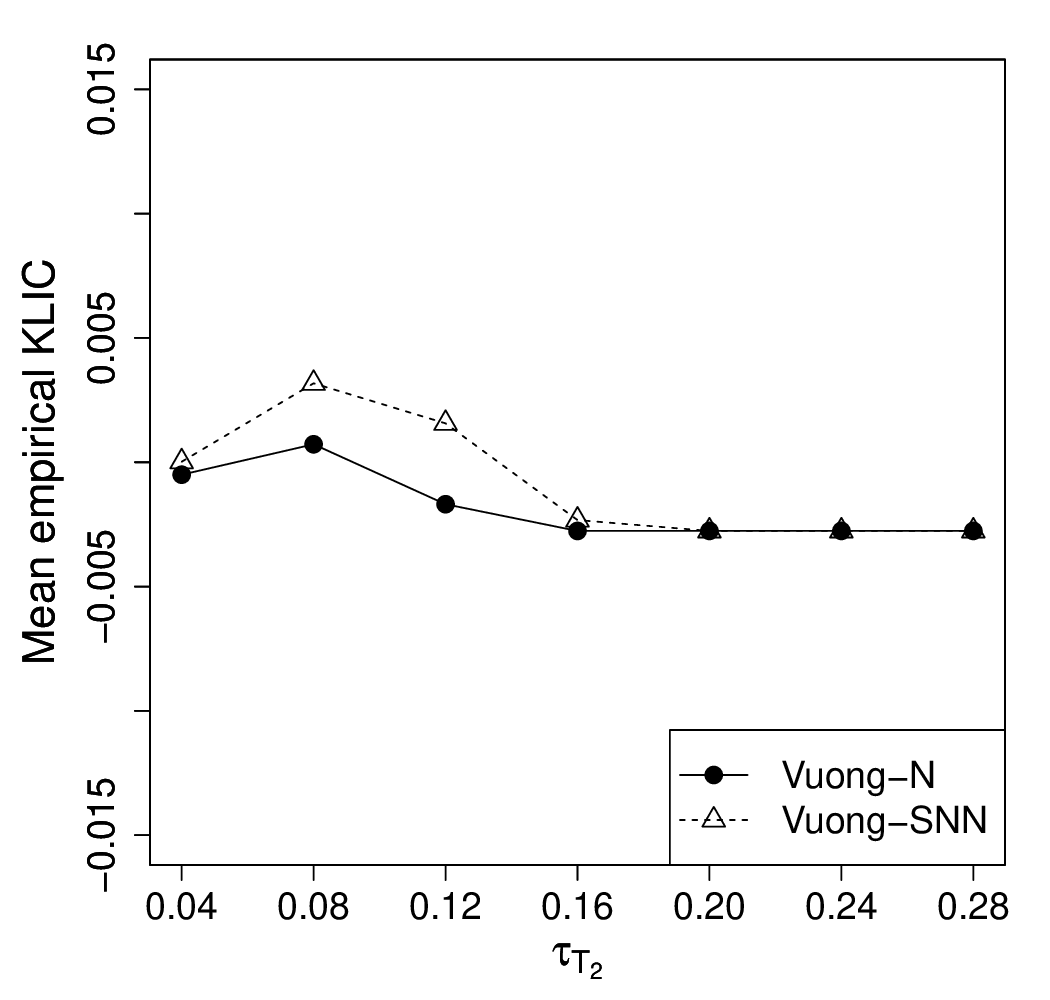}
        \caption{Mean empirical KLIC of the models chosen by Vuong test $(\alpha=0.05)$ with respect to $h_0$ for \(\tau_{T_1} = 0.20\) (top) and \(\tau_{T_1} = 0.28\) (bottom) with \(n = 200\) (left) and \(n = 500\) (right), corresponding to the conditions in Fig. \ref{fig:3d_num}.}
        \label{fig:3d_KLIC}
    \end{center}
\end{figure}

To further assess the performance of our tests, we computed the empirical Kullback-Leibler information criterion (KLIC) for the true model. This criterion was averaged over all repetitions:

\[
\overline{KLIC}(h_0) := \frac{1}{R} \sum_{r=1}^R \widehat{KLIC}(h_0, f_r, \hat{\bm\theta}_r),
\]
where \(f_r\) is the density of the approximating vine model in the \(r\)-th repetition with the estimated parameter \(\hat{\bm\theta}_r\). We selected either \(\bm F_{\bm\theta}\) or \(\bm G_{\bm\gamma}\) for each \(f_r\) based on the test results with \(\alpha = 0.05\). For each repetition \(r = 1, \ldots, R\), the empirical KLIC is defined as:

\[
\widehat{KLIC}(h_0, f_r, \hat{\bm\theta}_r) := \frac{1}{n} \sum_{i=1}^n \log \left[ h_0\left( \bm u_i^{0(r)} \right) \right] - \frac{1}{n} \sum_{i=1}^n \log \left[ f_r\left( \bm u_i^{0(r)} | \hat{\bm\theta}_r \right) \right],
\]
where \(\bm u_i^{0(r)}\) represents the \(i\)-th observation drawn from the true vine model \(h_0\). Note that the empirical KLIC may be negative, contrasting with its theoretical counterpart (see \citet[supplementary material]{Brechmann2012}). The results are shown in Fig. \ref{fig:3d_KLIC}. For \(\tau_{T_2} = 0.04\), the KLIC values under both Vuong-N and Vuong-SNN tests were nearly identical, indicating that both tests were equivalent in selecting models close to the true structure. When \(\tau_{T_2}\) increased to around \(0.12\) or \(0.16\), the KLIC values under Vuong-N test became smaller, showing better alignment with the true model. This difference was more pronounced for \(\tau_{T_1} = 0.20\) than for \(\tau_{T_1} = 0.28\). As \(\tau_{T_2}\) increased further, KLIC values under both Vuong-N and Vuong-SNN tests converged, indicating equivalent model selection performance.

\begin{figure}[tbh]
    \begin{center}
        \includegraphics[height=5.0cm]{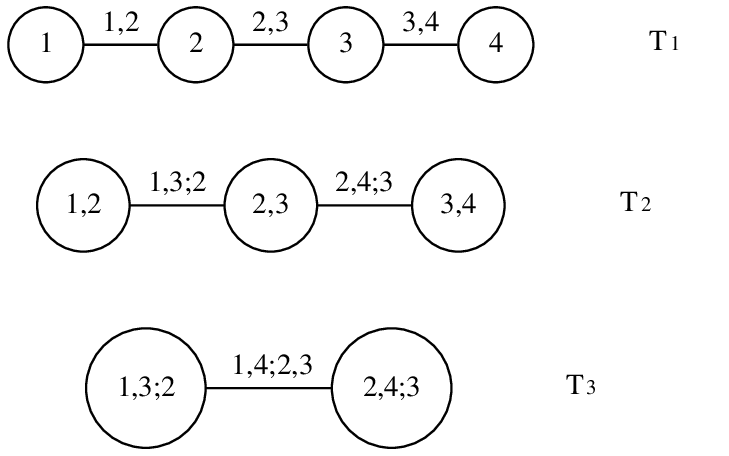}
        \caption{Four-dimensional R-vine structure used in simulation studies.}
        \label{fig:4d_D}
    \end{center}
\end{figure}

We extended our simulation study to a four-dimensional case to further assess the effectiveness of model selection tests for more complex vine copula structures. The structure of the four-dimensional R-vine model used in these simulations is shown in Fig. \ref{fig:4d_D}.

\begin{itemize}
    \item \textbf{True Models \((h_0)\):} Gaussian copulas were specified for each tree \(T_1\), \(T_2\), and \(T_3\), with Kendall's \(\tau\) values \(\tau_{T_1}\), \(\tau_{T_2}\), and \(\tau_{T_3}\) varying across different levels of dependence. Specifically, \(\tau_{T_1}\), \(\tau_{T_2}\), and \(\tau_{T_3}\) ranged from \(0.04\) to \(0.28\) in increments of \(0.04\), producing 343 unique configurations of true models.
    \item \textbf{Full Models \((\bm F_{\bm\theta})\):} Each copula in the structure, including \(c_{1,2}\), \(c_{2,3}\), \(c_{3,4}\), \(c_{1,3;2}\), \(c_{2,4;3}\), and \(c_{1,4;2,3}\), was specified as Gaussian to allow for full flexibility.
    \item \textbf{Truncated Models 1 \((\bm G^1_{\bm\gamma_1})\):} The copulas \(c_{1,2}\), \(c_{2,3}\), and \(c_{3,4}\) were set as Gaussian, while the copulas \(c_{1,3;2}\), \(c_{2,4;3}\), and \(c_{1,4;2,3}\) were modeled as independence copulas.
    \item \textbf{Truncated Models 2 \((\bm G^2_{\bm\gamma_2})\):} The copulas \(c_{1,2}\), \(c_{2,3}\), \(c_{3,4}\), \(c_{1,3;2}\), and \(c_{2,4;3}\) were Gaussian, while the copula \(c_{1,4;2,3}\) was modeled as an independence copula.
\end{itemize}

As in the three-dimensional case, we simulated \(n \in \{100, 200, 500, 1000\}\) observations from each true model, optimizing both the full model \(\bm F_{\bm\theta}\) and the two truncated models \(\bm G^1_{\bm\gamma_1}\) and \(\bm G^2_{\bm\gamma_2}\) to the simulated data. We then conducted analyses similar to the three-dimensional case, comparing two model pairs: \(\bm G^1_{\bm\gamma_1}\) versus \(\bm G^2_{\bm\gamma_2}\), and \(\bm G^2_{\bm\gamma_2}\) versus \(\bm F_{\bm\theta}\). Each scenario was again repeated $R=300$ times.

\begin{figure}[tbh]
    \begin{center}
        \includegraphics[height=6cm]{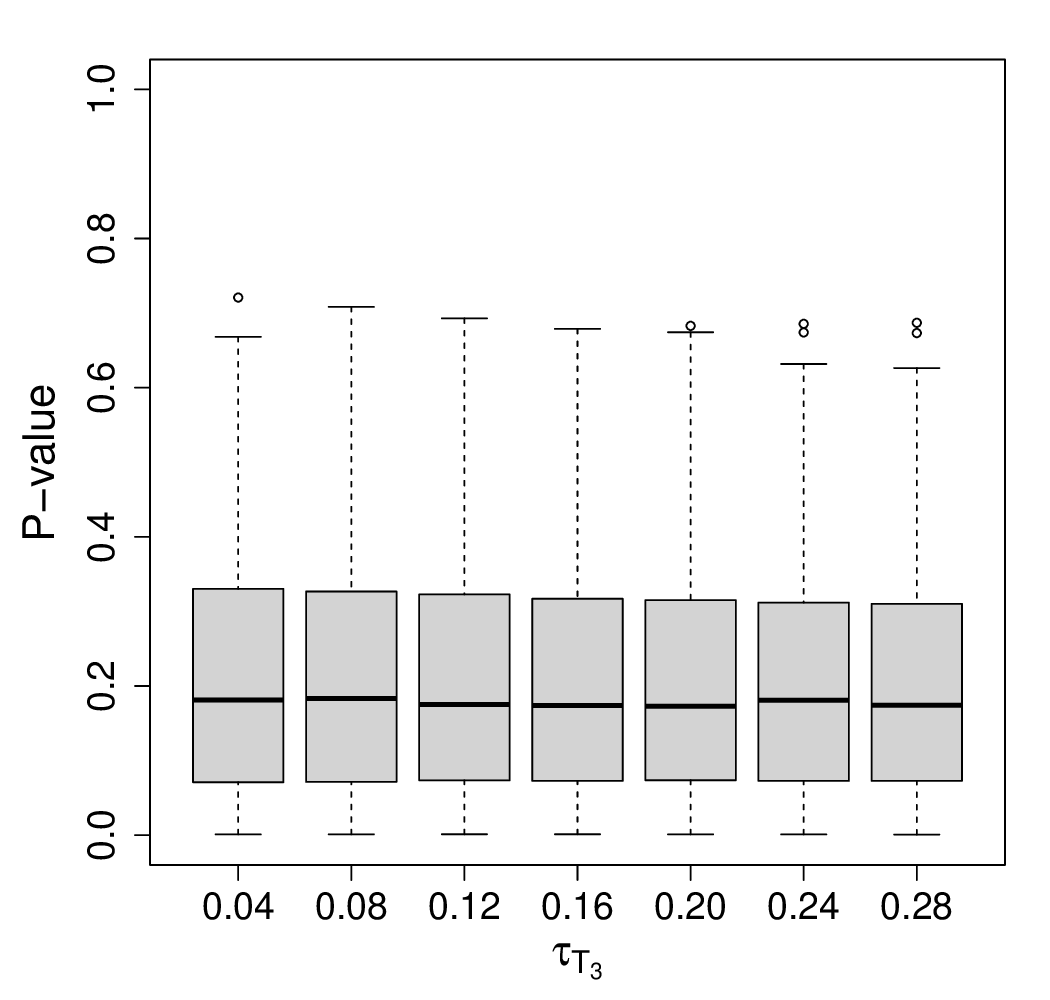}
        \includegraphics[height=6cm]{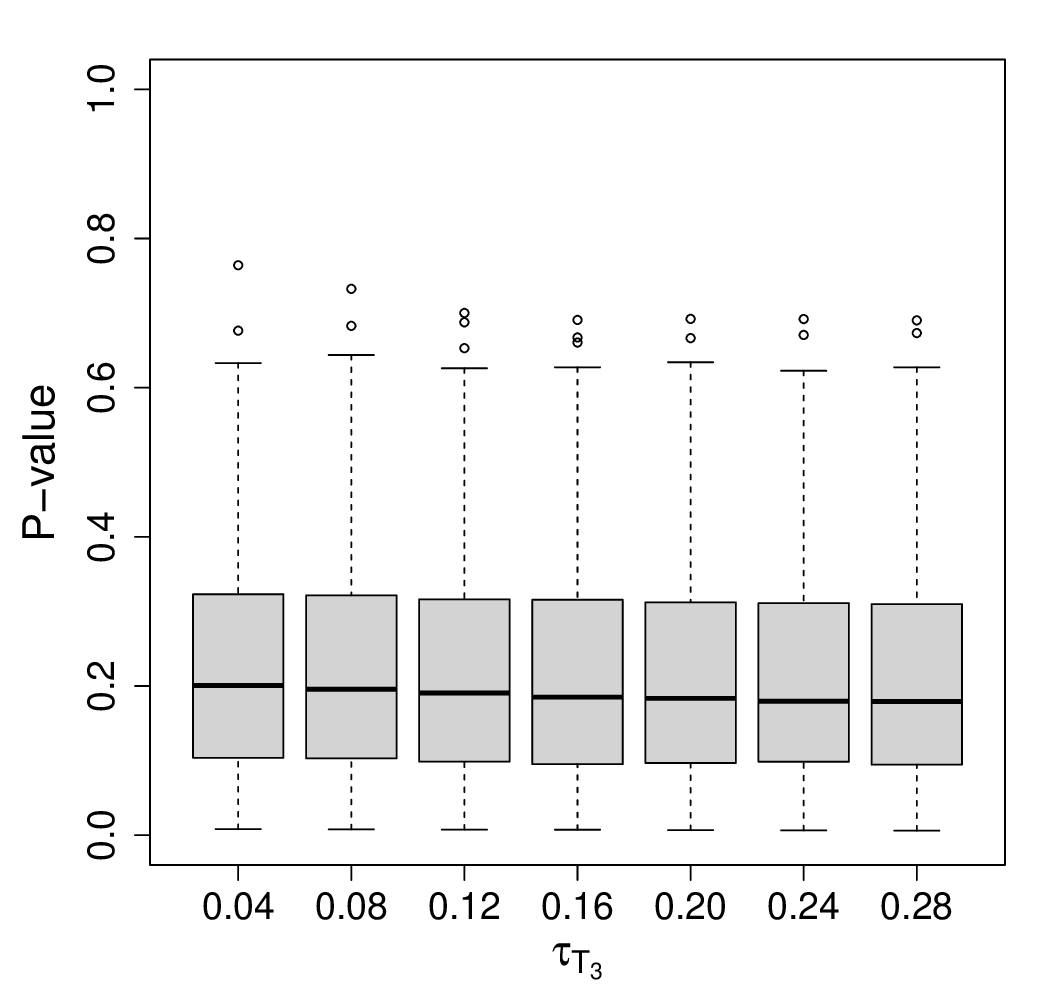}\\
        \includegraphics[height=6cm]{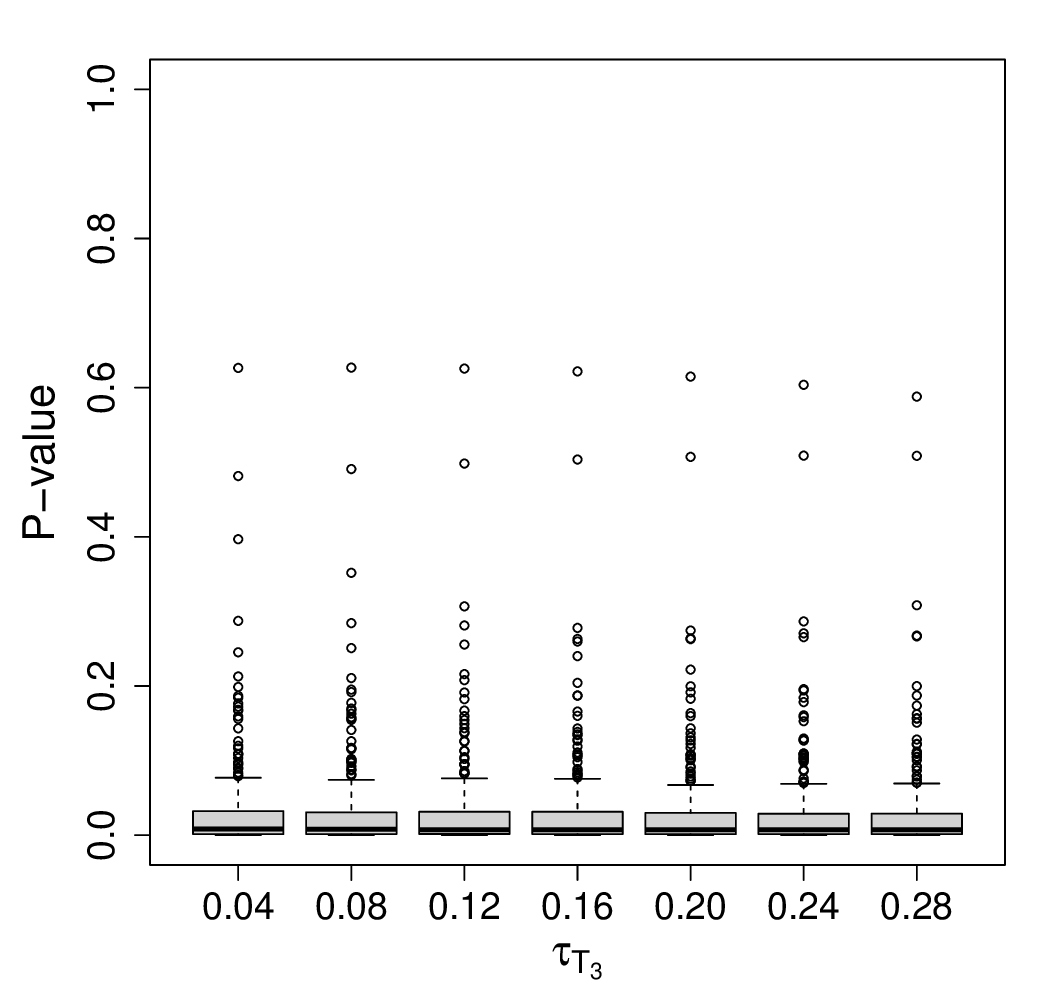}
        \includegraphics[height=6cm]{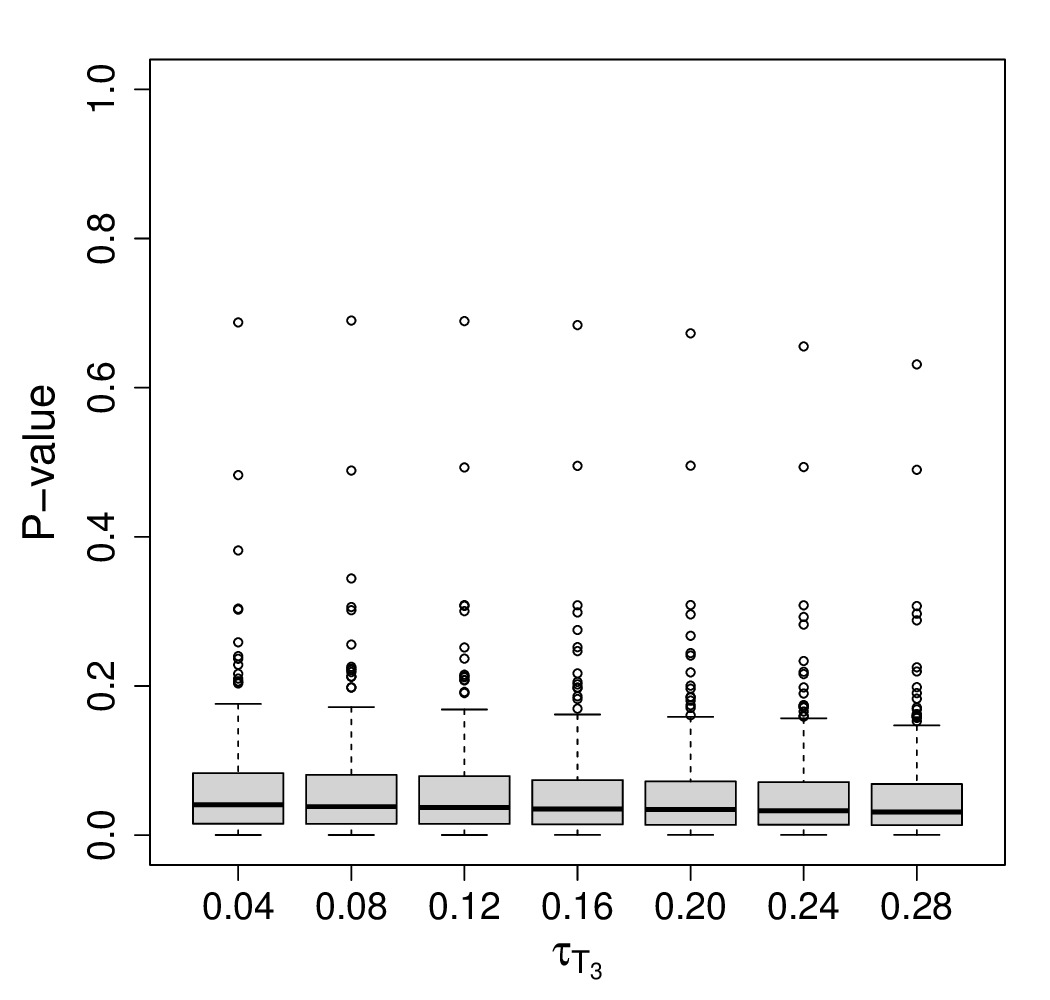}
        \caption{P-values obtained by Vuong-N (left) and Vuong-SNN (right) tests comparing \(\bm G^1_{\bm\gamma_1}\) and \(\bm G^2_{\bm\gamma_2}\) for \(\tau_{T_1} = 0.12\) and \(\tau_{T_2} = 0.08\) with \(n = 200\) (top) and \(n = 500\) (bottom). The corresponding \(\tau_{T_3}\) values are as shown.}
        \label{fig:4d_D_tr1_pval1}
    \end{center}
\end{figure}

\begin{figure}[tbh]
    \begin{center}
        \includegraphics[height=6cm]{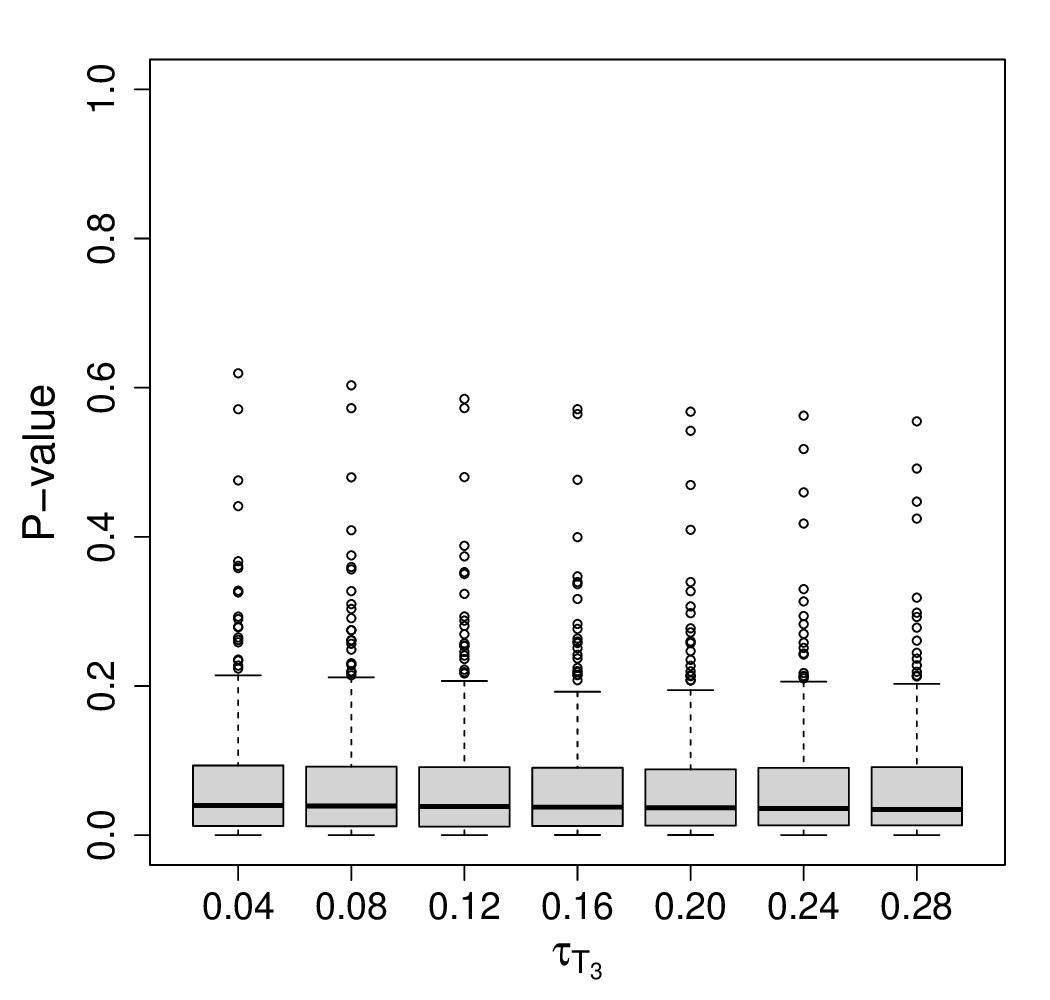}
        \includegraphics[height=6cm]{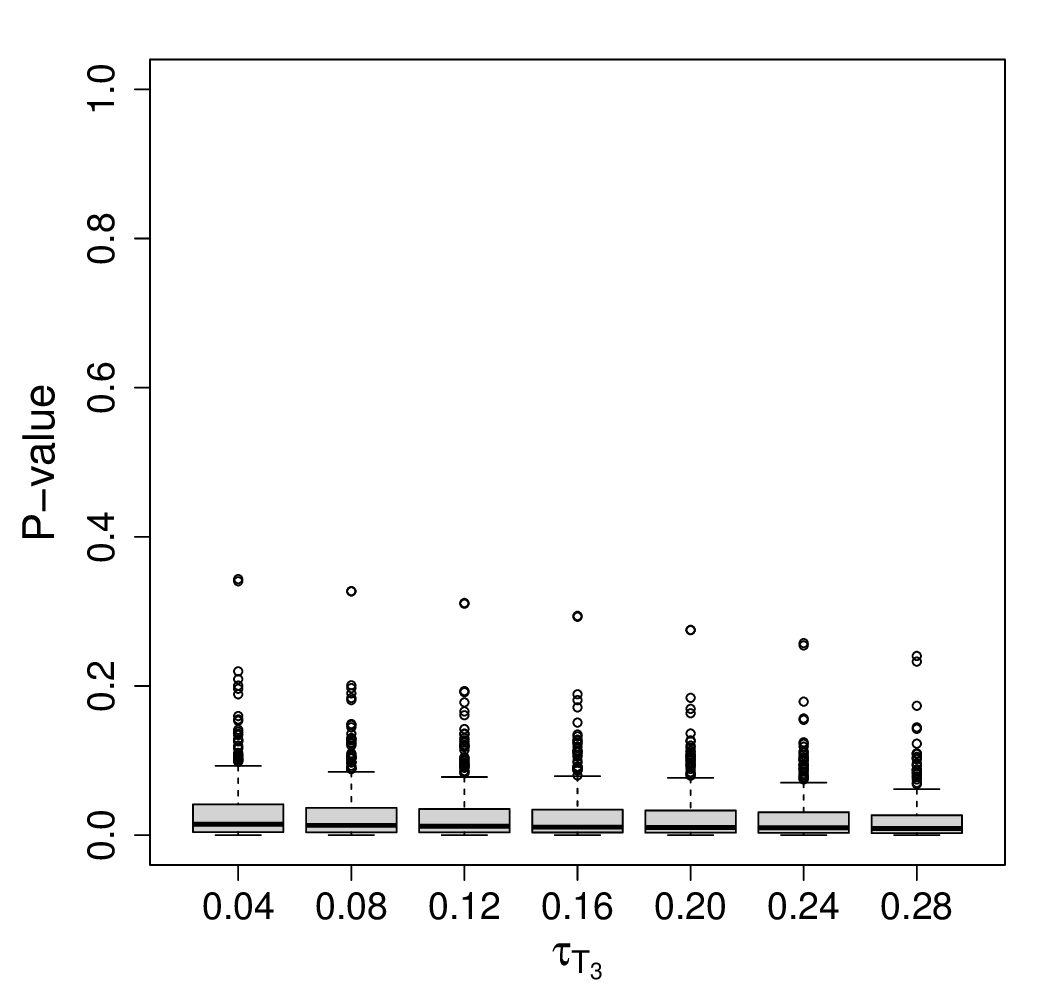}\\
        \includegraphics[height=6cm]{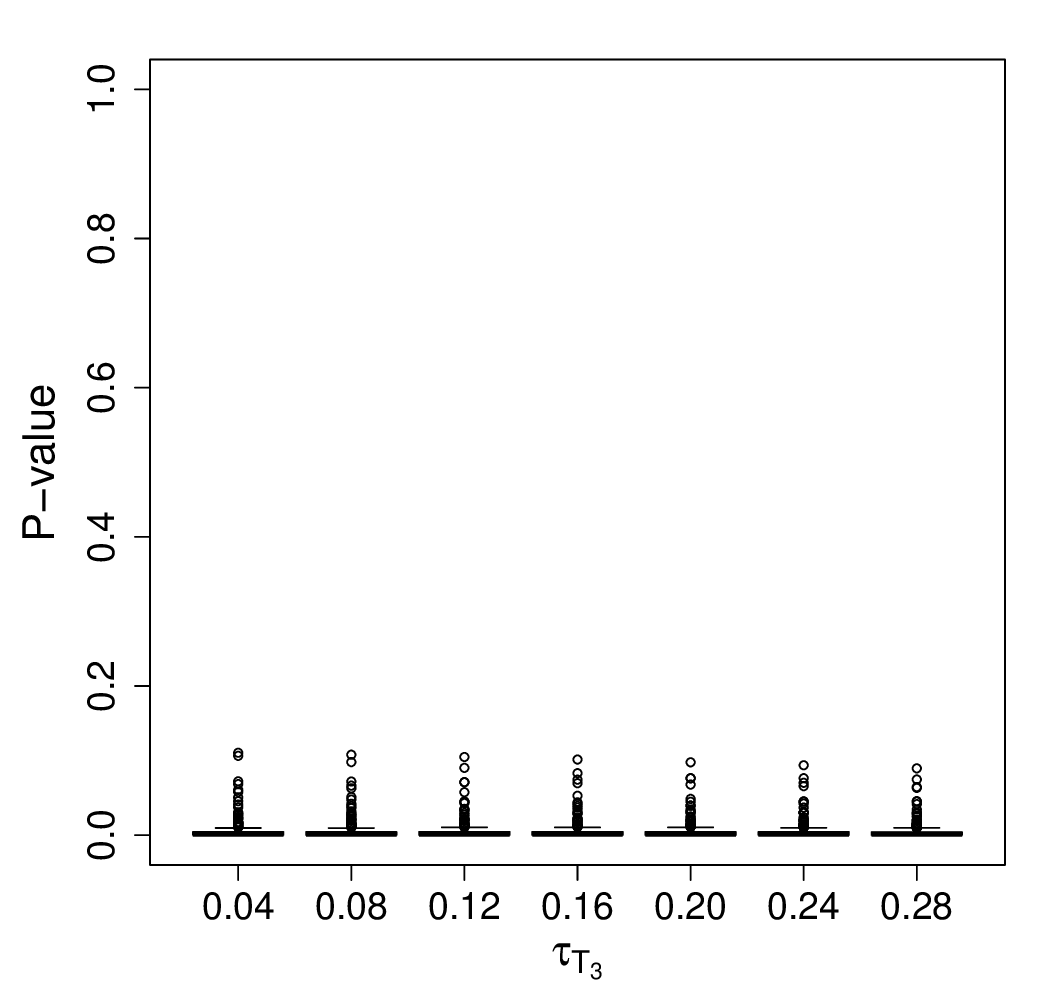}
        \includegraphics[height=6cm]{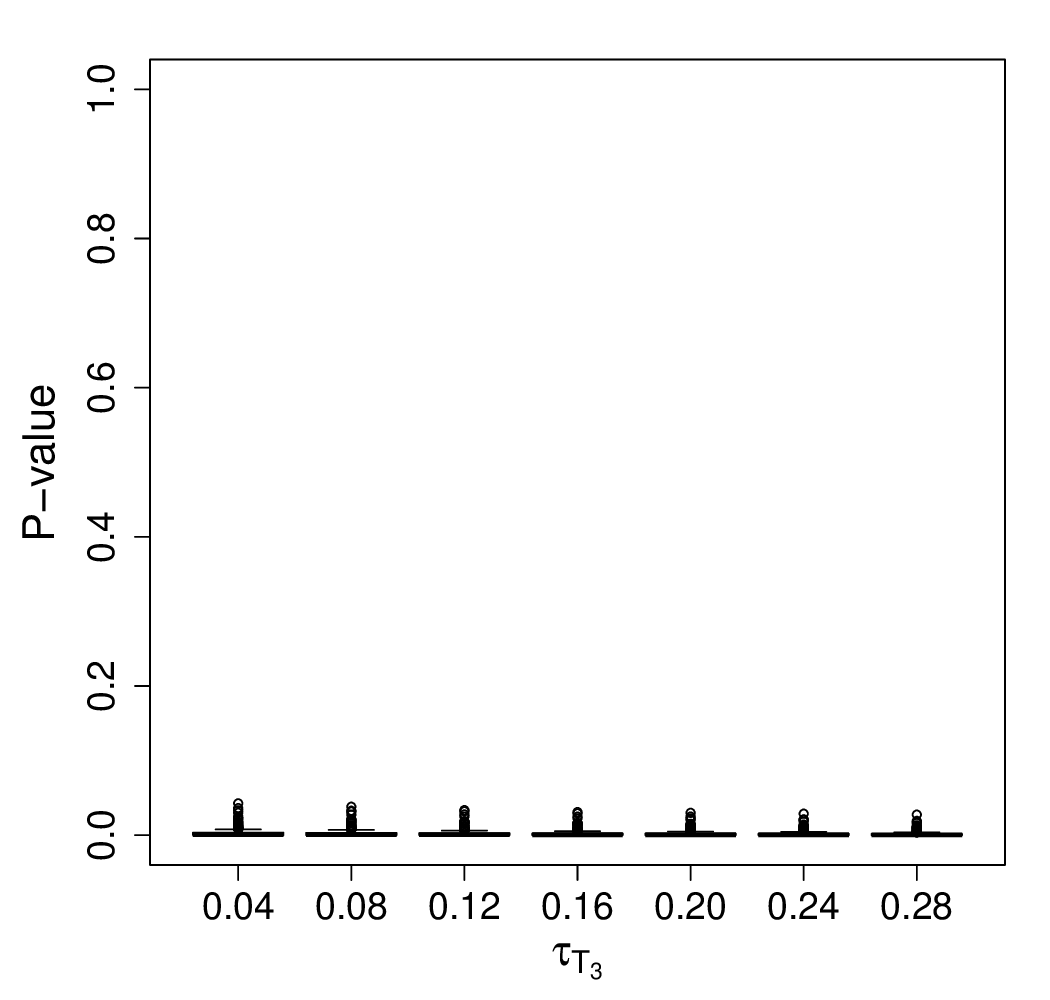}
        \caption{P-values obtained by Vuong-N (left) and Vuong-SNN (right) tests comparing \(\bm G^1_{\bm\gamma_1}\) and \(\bm G^2_{\bm\gamma_2}\) for \(\tau_{T_1} = 0.28\) and \(\tau_{T_2} = 0.20\) with \(n = 100\) (top) and \(n = 200\) (bottom). The corresponding \(\tau_{T_3}\) values are as shown.}
        \label{fig:4d_D_tr1_pval2}
    \end{center}
\end{figure}

\begin{figure}[tbh]
    \begin{center}
        \includegraphics[height=6cm]{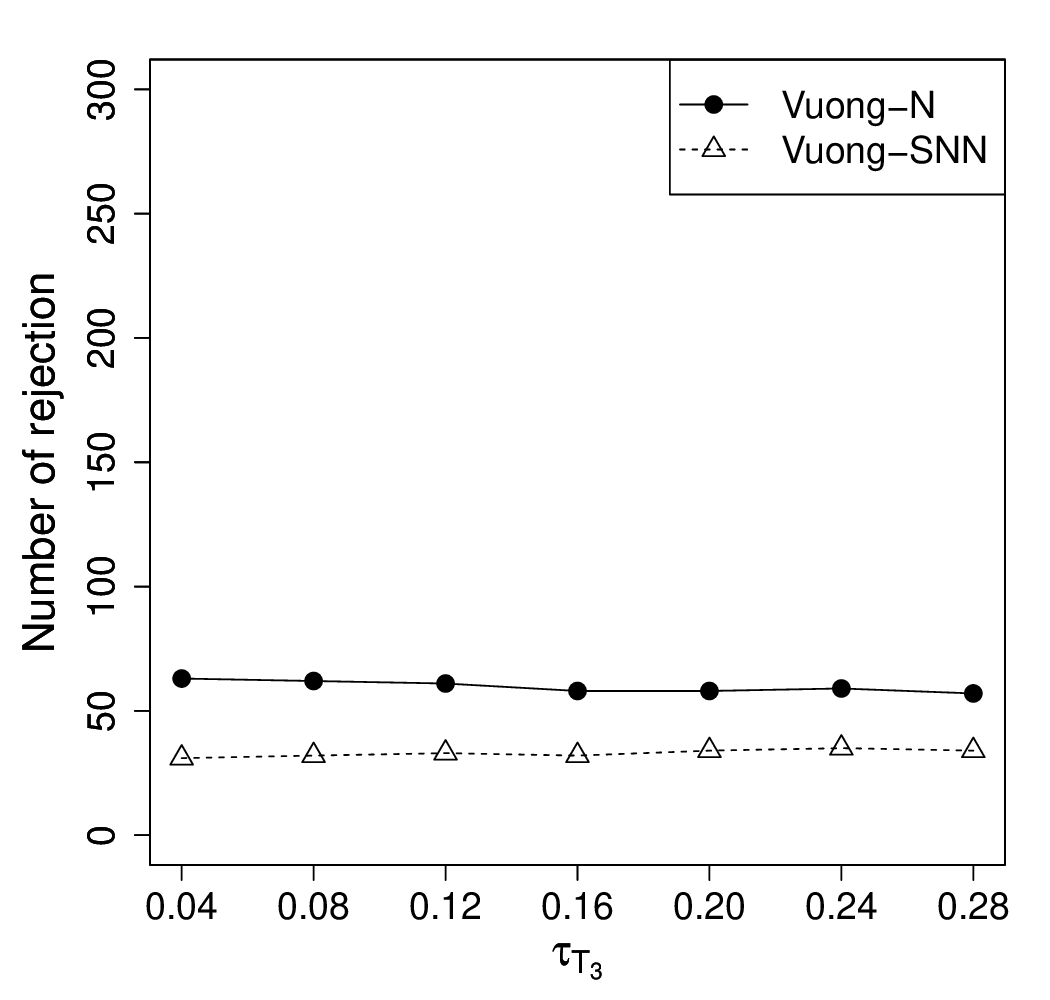}
        \includegraphics[height=6cm]{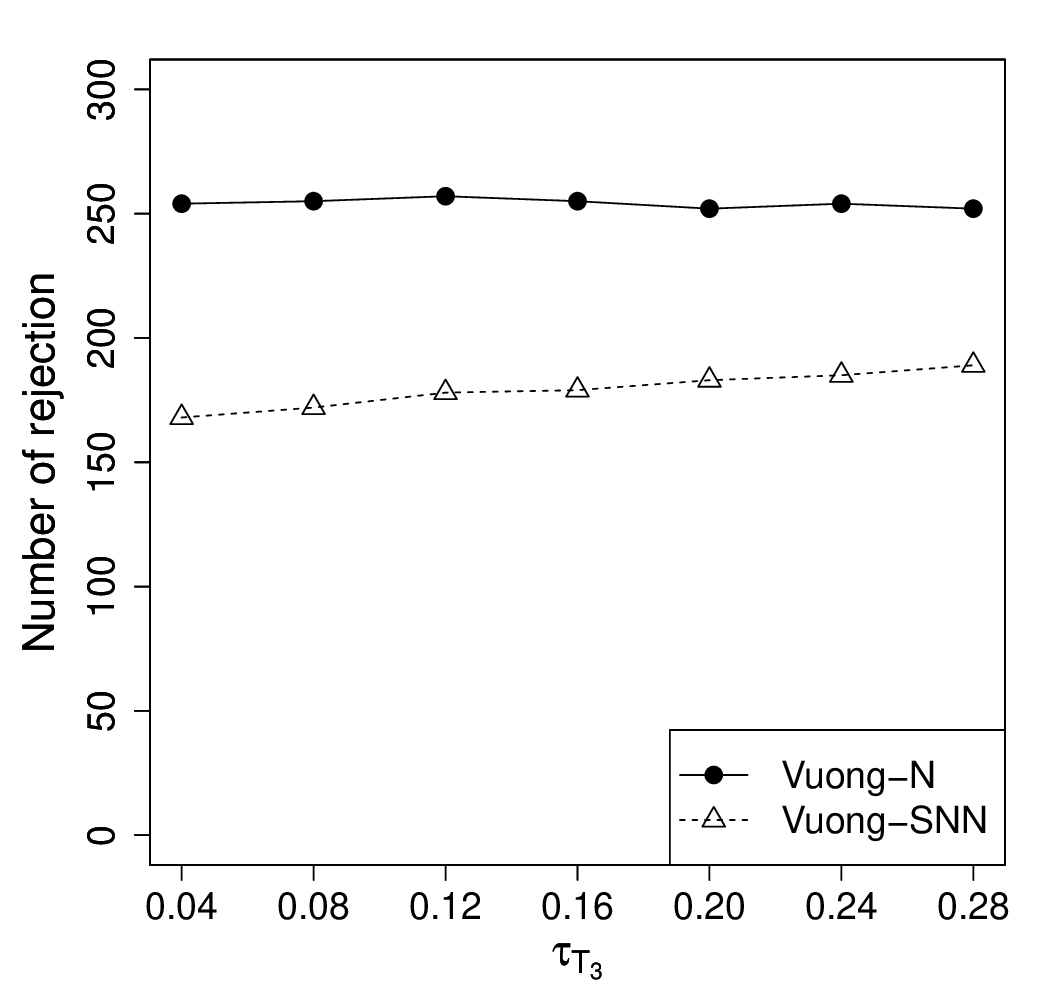}\\
        \includegraphics[height=6cm]{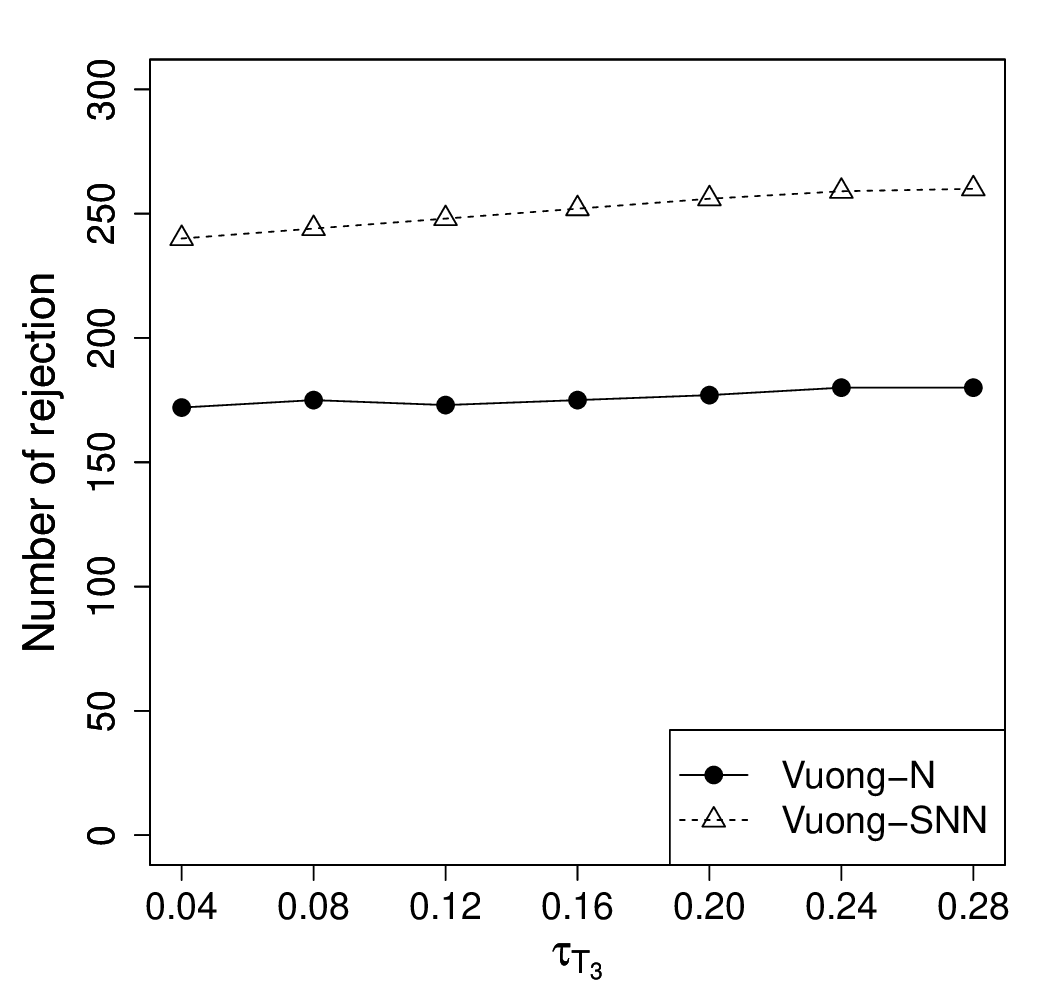}
        \includegraphics[height=6cm]{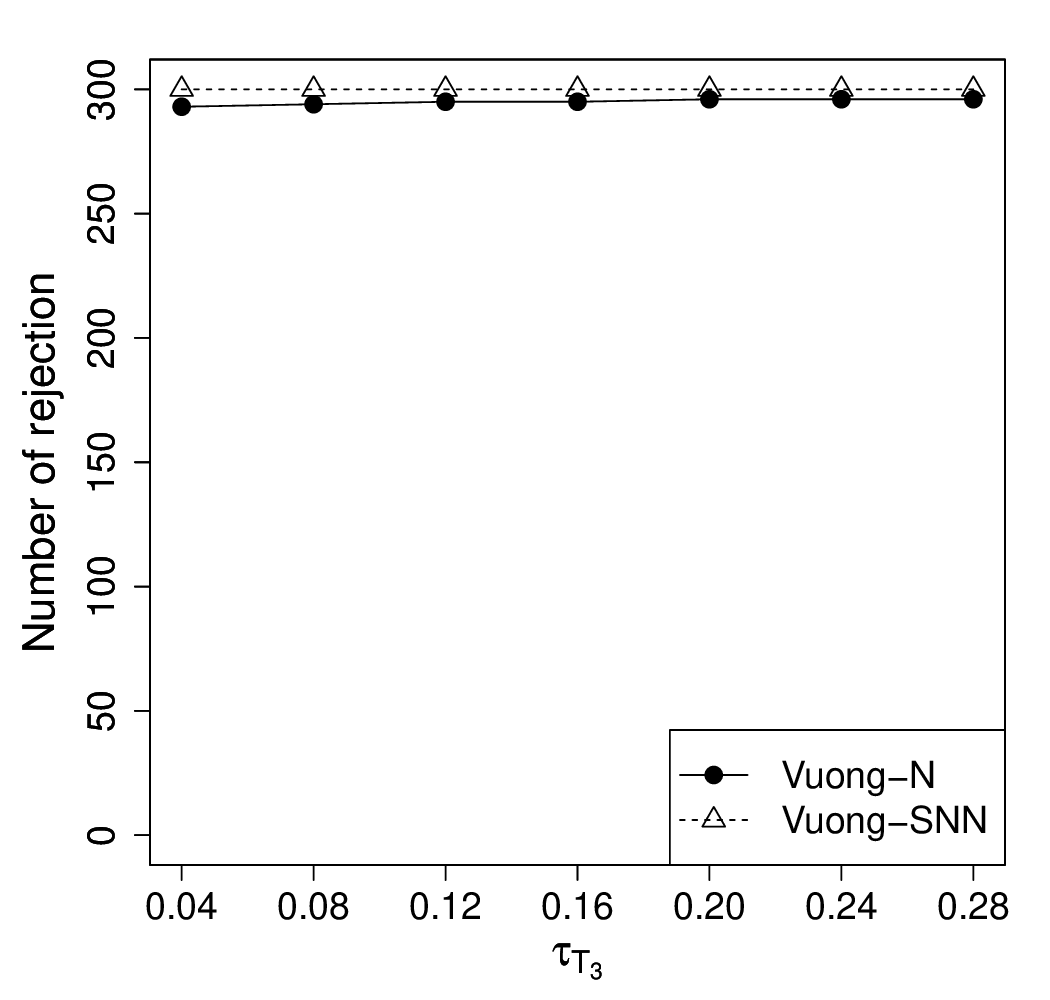}
        \caption{The number of rejections by the Vuong test $(\alpha=0.05)$ comparing \(\bm G^1_{\bm\gamma_1}\) and \(\bm G^2_{\bm\gamma_2}\) for \(\tau_{T_1} = 0.12\) and \(\tau_{T_2} = 0.08\) with \(n = 200\) (top left) and \(n = 500\) (top right) and \(\tau_{T_1} = 0.28\) and \(\tau_{T_2} = 0.20\) with \(n = 100\) (bottom left) and \(n = 200\) (bottom right), corresponding to the conditions of Figs. \ref{fig:4d_D_tr1_pval1} and \ref{fig:4d_D_tr1_pval2}.}
        \label{fig:4d_D_tr1_num}
    \end{center}
\end{figure}

\begin{figure}[tbh]
    \begin{center}
        \includegraphics[height=6cm]{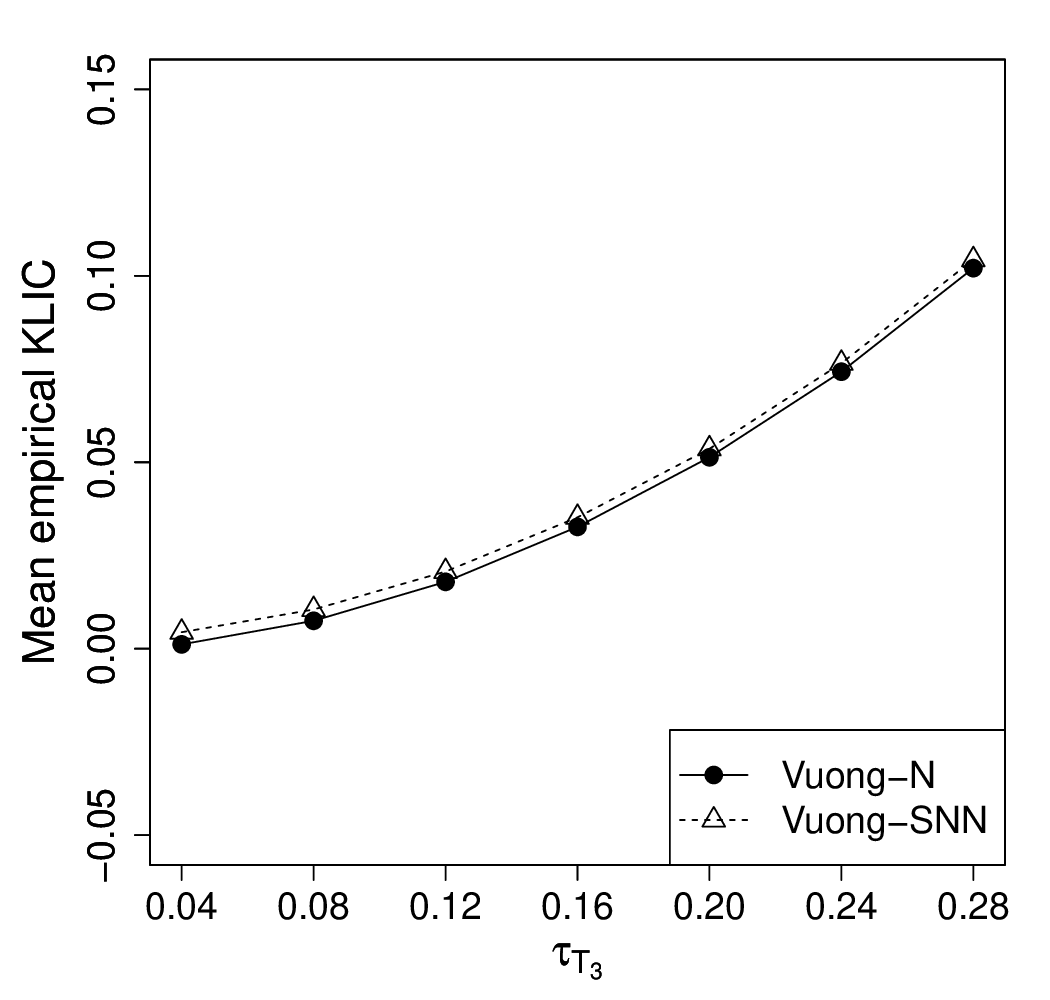}
        \includegraphics[height=6cm]{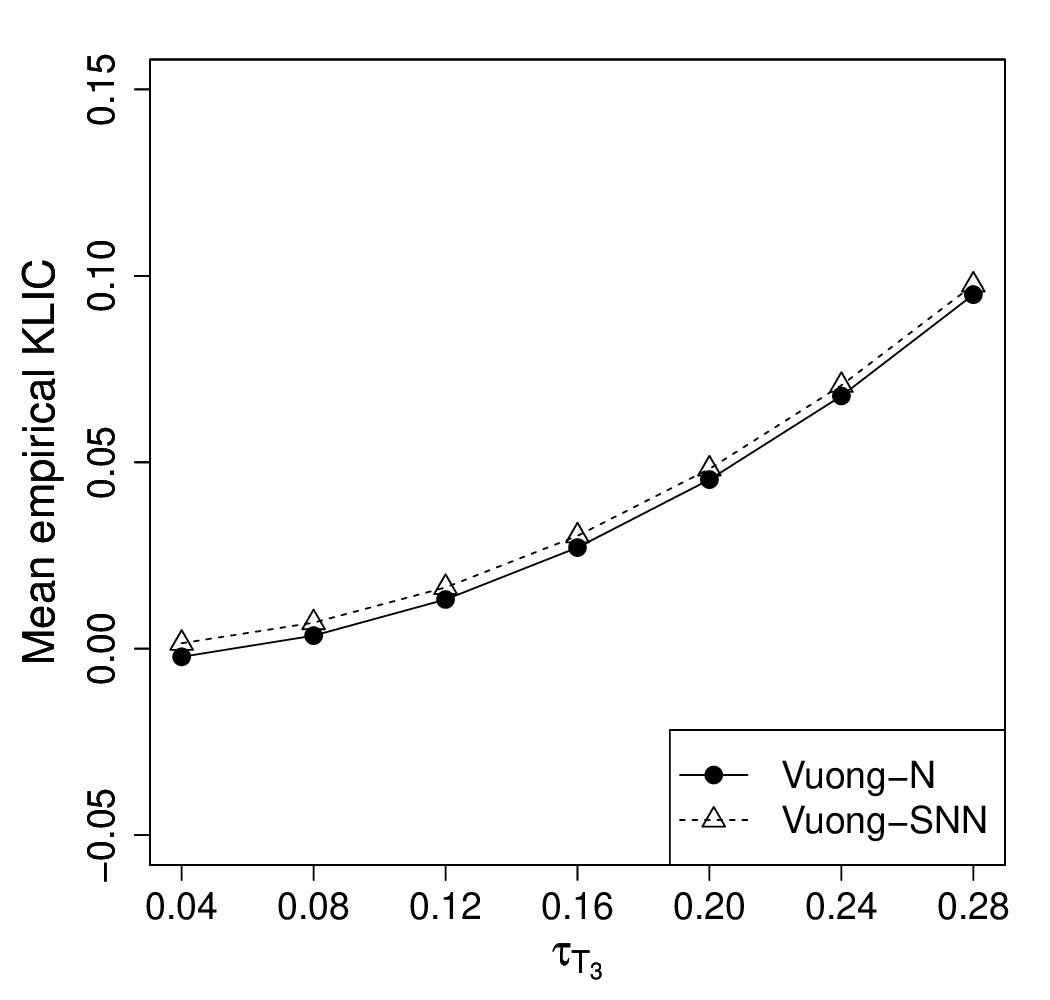}\\
        \includegraphics[height=6cm]{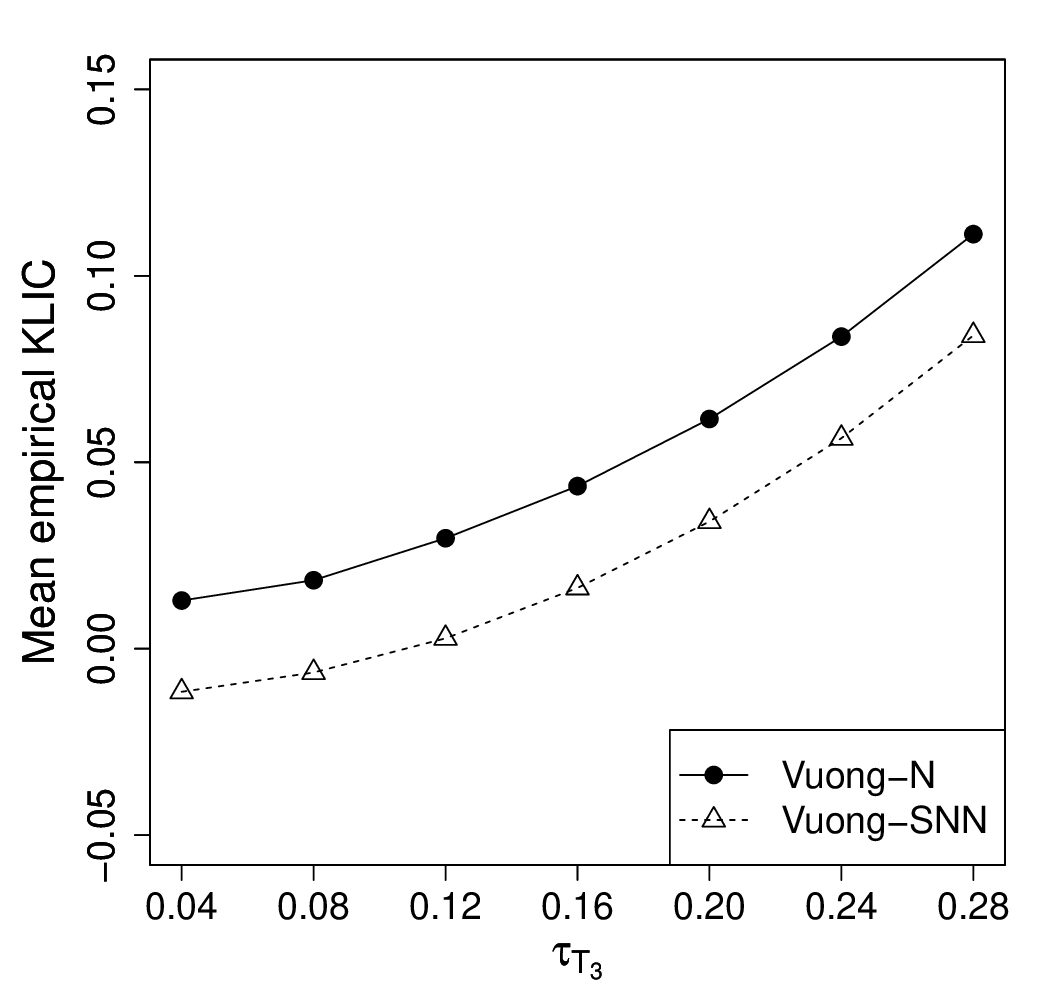}
        \includegraphics[height=6cm]{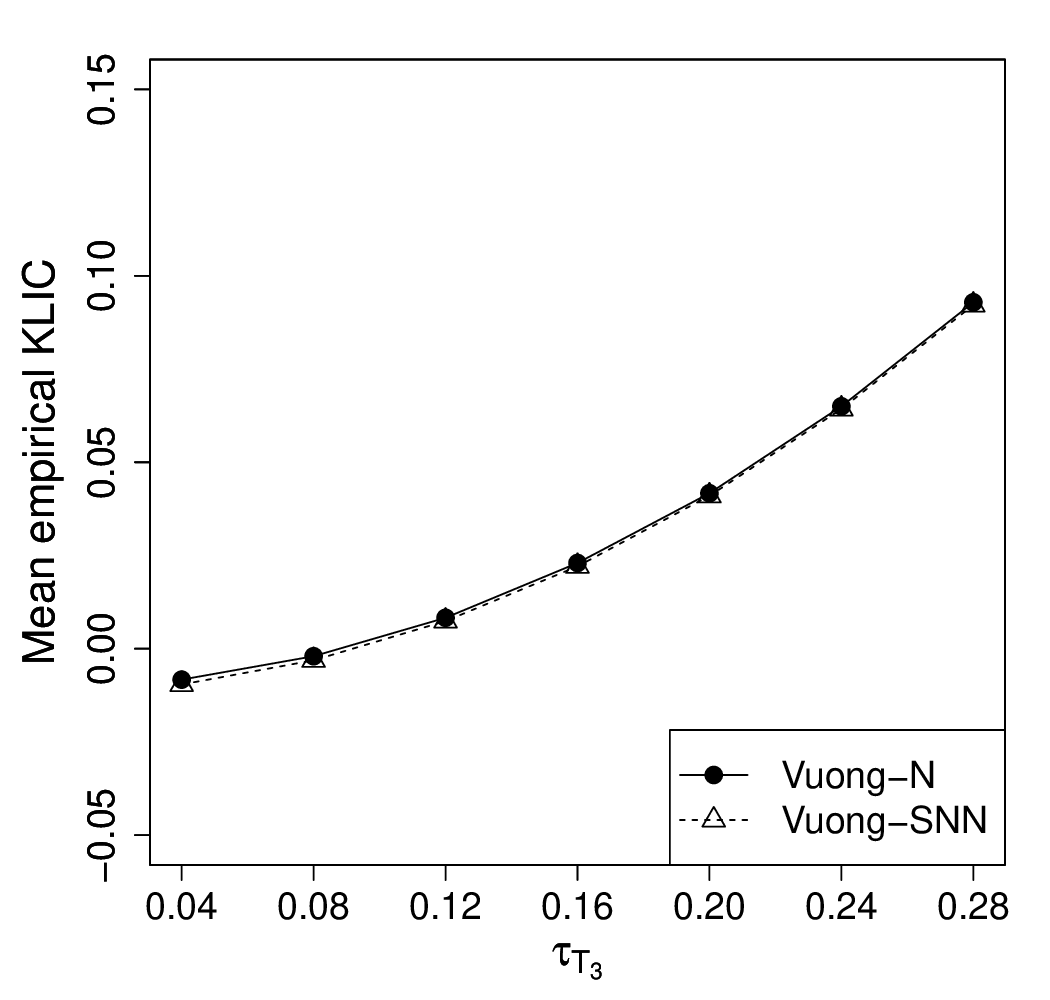}
        \caption{Mean empirical KLIC of the models chosen by Vuong test $(\alpha=0.05)$ comparing \(\bm G^1_{\bm\gamma_1}\) and \(\bm G^2_{\bm\gamma_2}\) with respect to $h_0$ for  \(\tau_{T_1} = 0.12\) and \(\tau_{T_2} = 0.08\) with \(n = 200\) (top left) and \(n = 500\) (top right) and \(\tau_{T_1} = 0.28\) and \(\tau_{T_2} = 0.20\) with \(n = 100\) (bottom left) and \(n = 200\) (bottom right), corresponding to the conditions of Fig. \ref{fig:4d_D_tr1_num}.}
        \label{fig:4d_D_tr1_KLIC}
    \end{center}
\end{figure}

The p-values from the Vuong-N and Vuong-SNN tests comparing \(\bm G^1_{\bm\gamma_1}\) and \(\bm G^2_{\bm\gamma_2}\) for \(\tau_{T_1} = 0.12\) and \(\tau_{T_2} = 0.08\) are shown in Fig. \ref{fig:4d_D_tr1_pval1}. Little variation in \(\tau_{T_3}\) was observed, which is expected since the corresponding pair copula was modeled as independent in both \(\bm G^1_{\bm\gamma_1}\) and \(\bm G^2_{\bm\gamma_2}\). In comparing Vuong-N and Vuong-SNN tests, the median of p-values obtained by Vuong-N test was slightly lower. However, for \(\tau_{T_1} = 0.28\) and \(\tau_{T_2} = 0.20\), the median of p-values obtained by Vuong-N test was higher than Vuong-SNN test, as shown in Fig. \ref{fig:4d_D_tr1_pval2}. This suggests that the condition \(f(\cdot|\cdot; \bm\theta_*) = g(\cdot|\cdot; \bm\gamma_*)\) may not hold well in cases of higher correlation. Figures \ref{fig:4d_D_tr1_num} and \ref{fig:4d_D_tr1_KLIC} show the corresponding rejection counts and mean empirical KLIC. The Vuong-N test achieved higher rejection rates in weaker correlation cases, while the Vuong-SNN test performed better in stronger correlation cases, which aligns with the p-value results. The mean empirical KLIC increased with \(\tau_{T_3}\), indicating that misspecification grew in both \(\bm G^1_{\bm\gamma_1}\) and \(\bm G^2_{\bm\gamma_2}\).


\begin{figure}[tbh]
    \begin{center}
        \includegraphics[height=6cm]{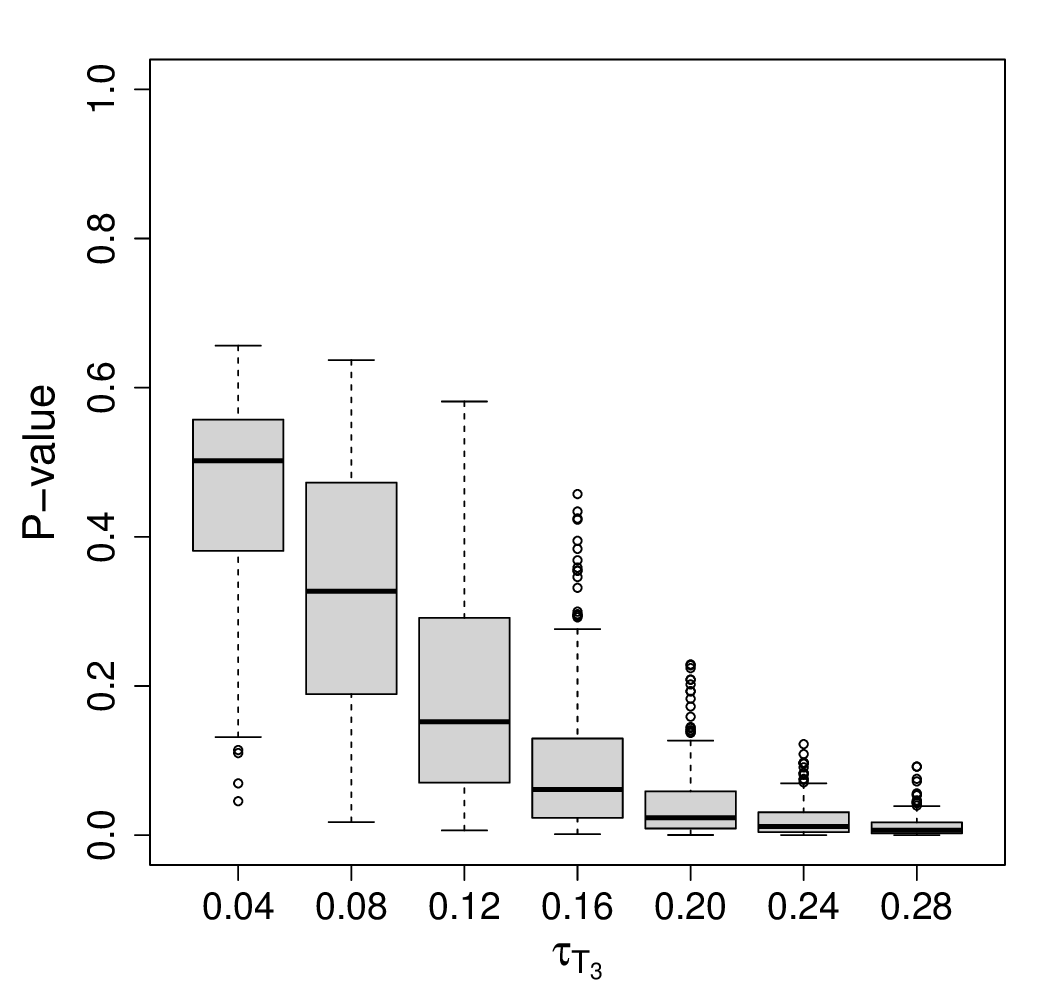}
        \includegraphics[height=6cm]{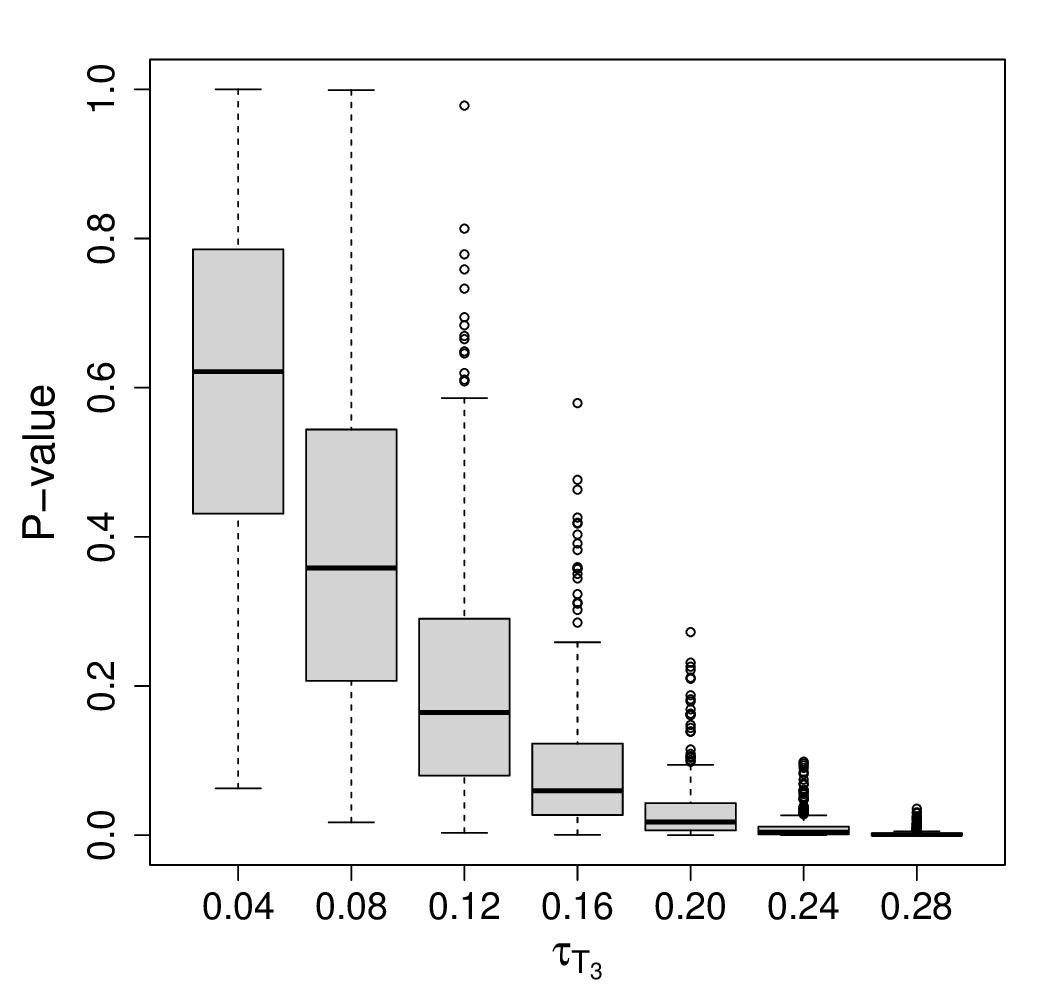}\\
        \includegraphics[height=6cm]{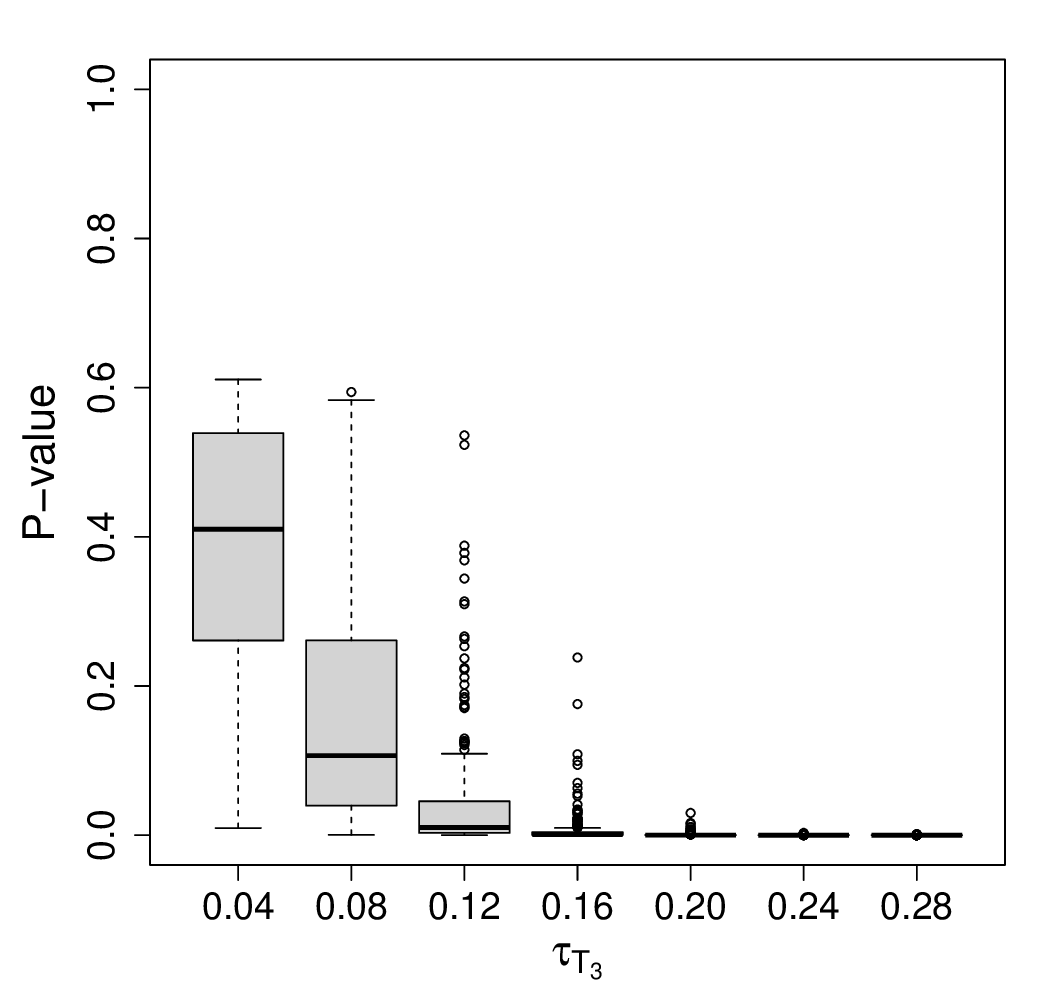}
        \includegraphics[height=6cm]{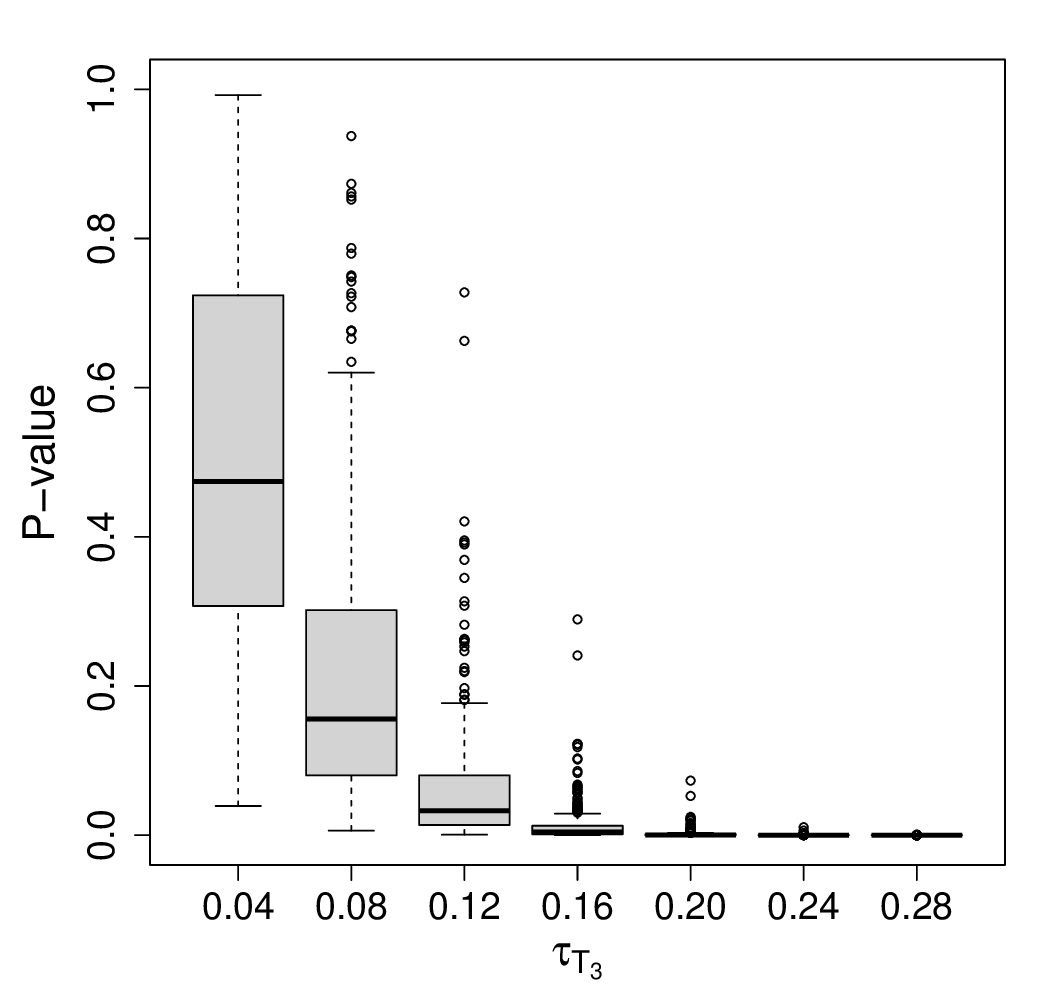}
        \caption{P-values obtained by Vuong-N (left) and Vuong-SNN (right) tests comparing \(\bm G^2_{\bm\gamma_2}\) and \(\bm F_{\bm\theta}\) for \(\tau_{T_1} = 0.12\) and \(\tau_{T_2} = 0.08\) with \(n = 200\) (top) and \(n = 500\) (bottom). The corresponding \(\tau_{T_3}\) values are as shown.}
        \label{fig:4d_D_tr2_pval1}
    \end{center}
\end{figure}

\begin{figure}[tbh]
    \begin{center}
        \includegraphics[height=6cm]{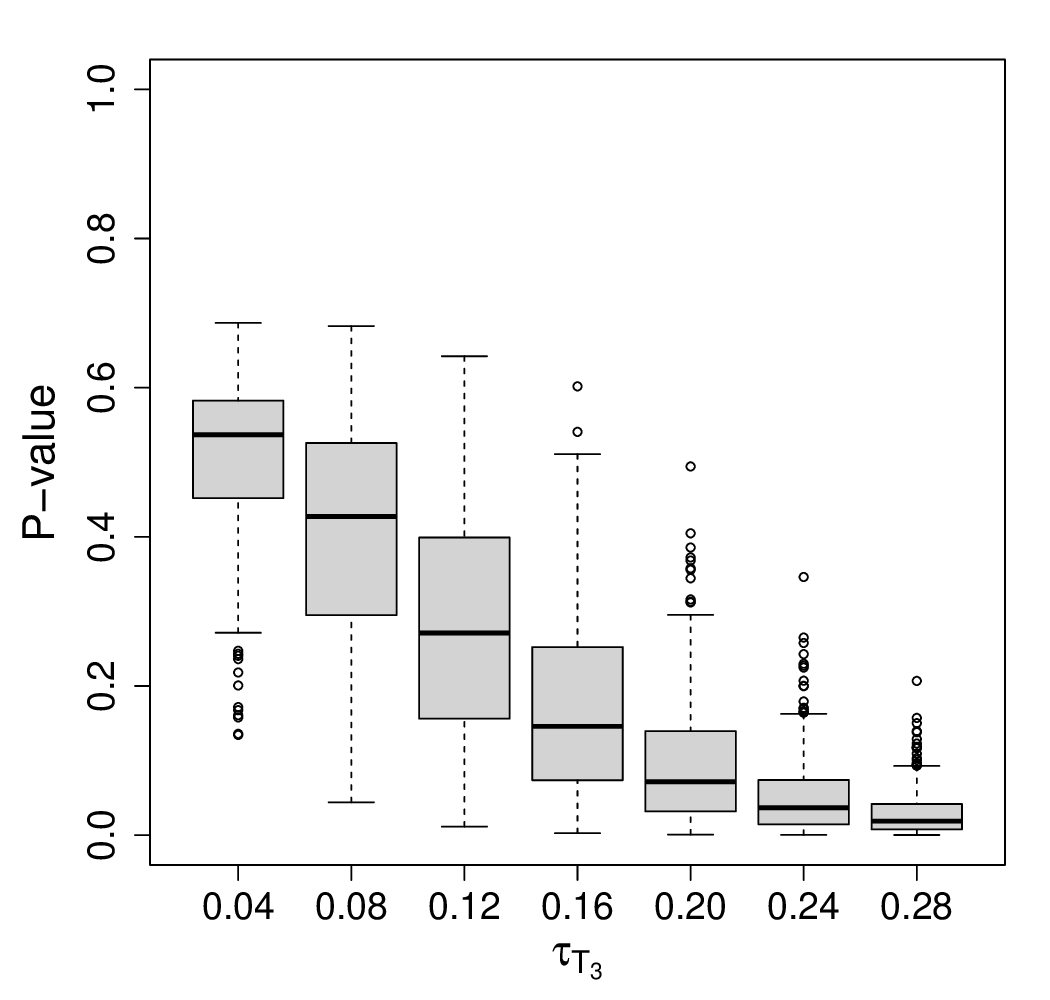}
        \includegraphics[height=6cm]{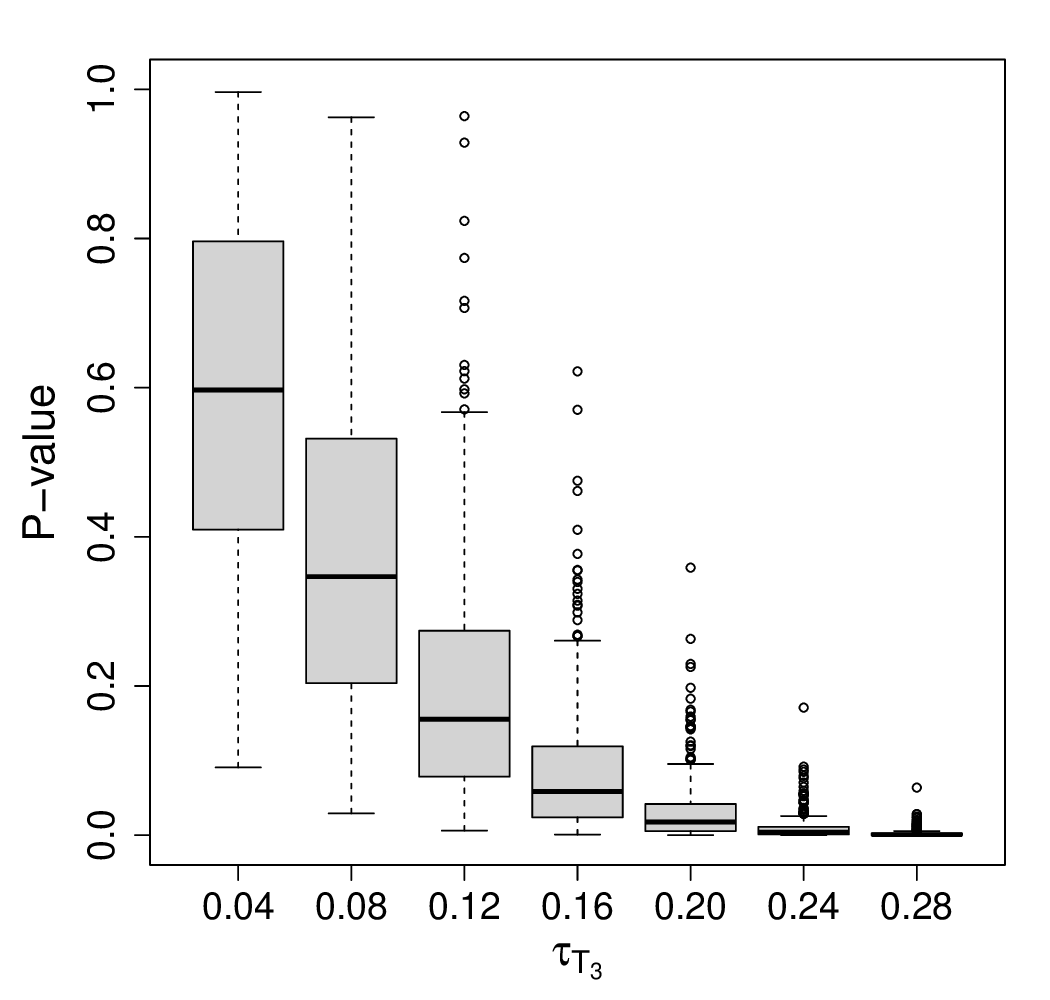}\\
        \includegraphics[height=6cm]{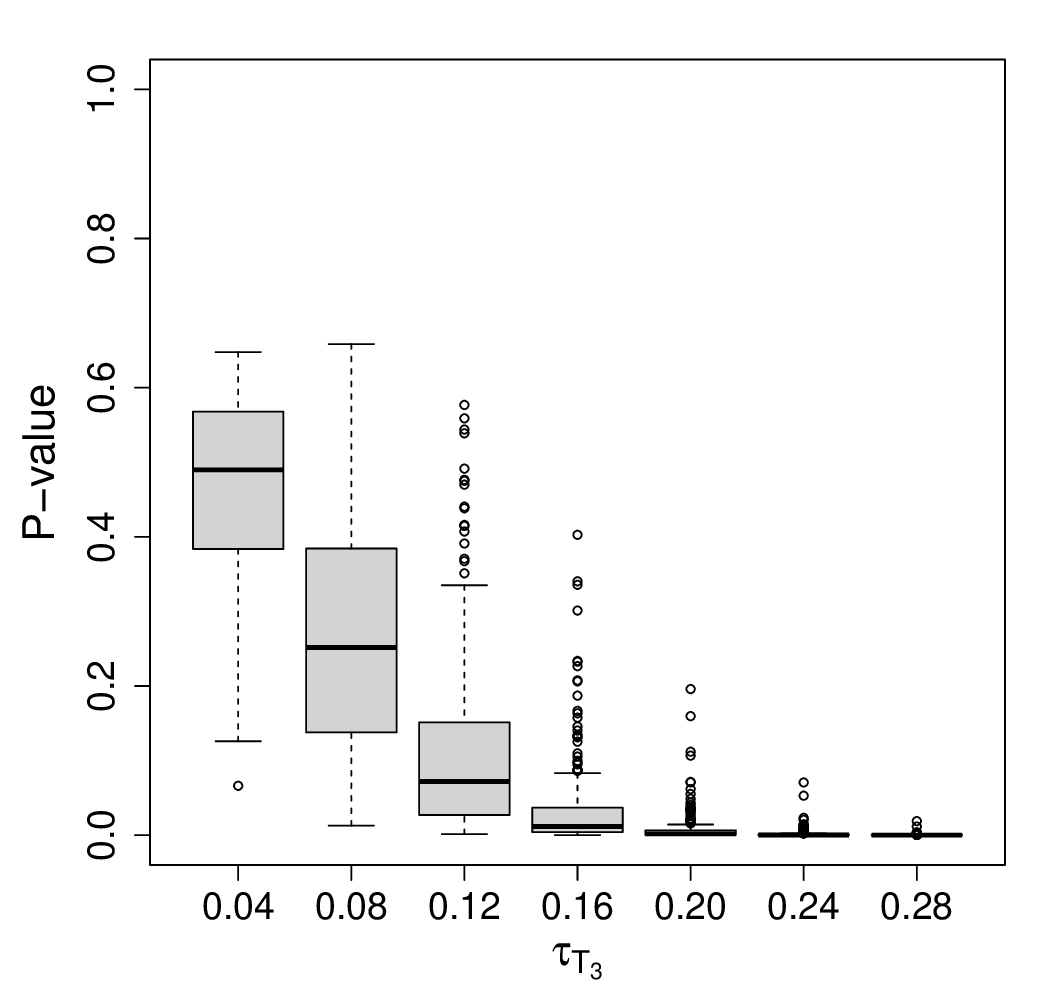}
        \includegraphics[height=6cm]{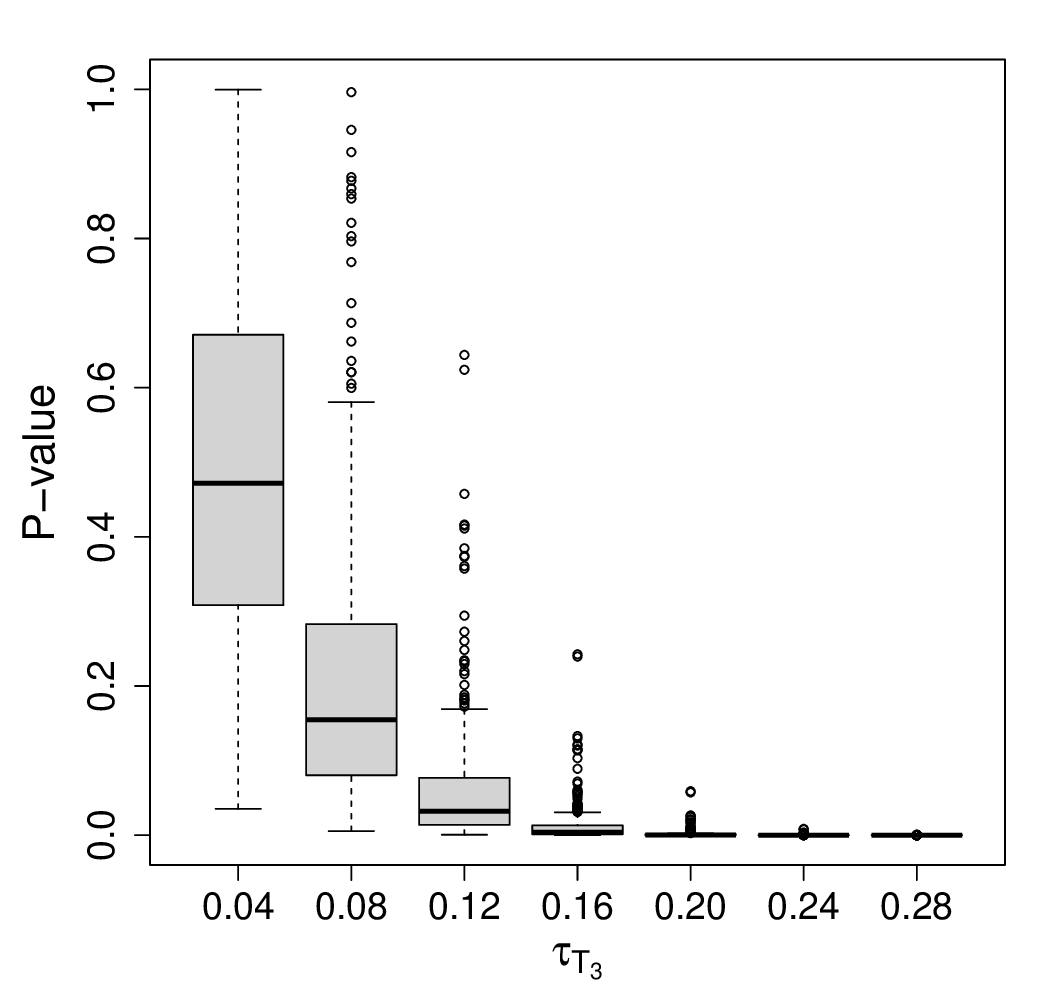}
        \caption{P-values obtained by Vuong-N (left) and Vuong-SNN (right) tests comparing \(\bm G^2_{\bm\gamma_2}\) and \(\bm F_{\bm\theta}\) for \(\tau_{T_1} = 0.28\) and \(\tau_{T_2} = 0.28\) with \(n = 200\) (top) and \(n = 500\) (bottom). The corresponding \(\tau_{T_3}\) values are as shown.}
        \label{fig:4d_D_tr2_pval2}
    \end{center}
\end{figure}

\begin{figure}[tbh]
    \begin{center}
        \includegraphics[height=6cm]{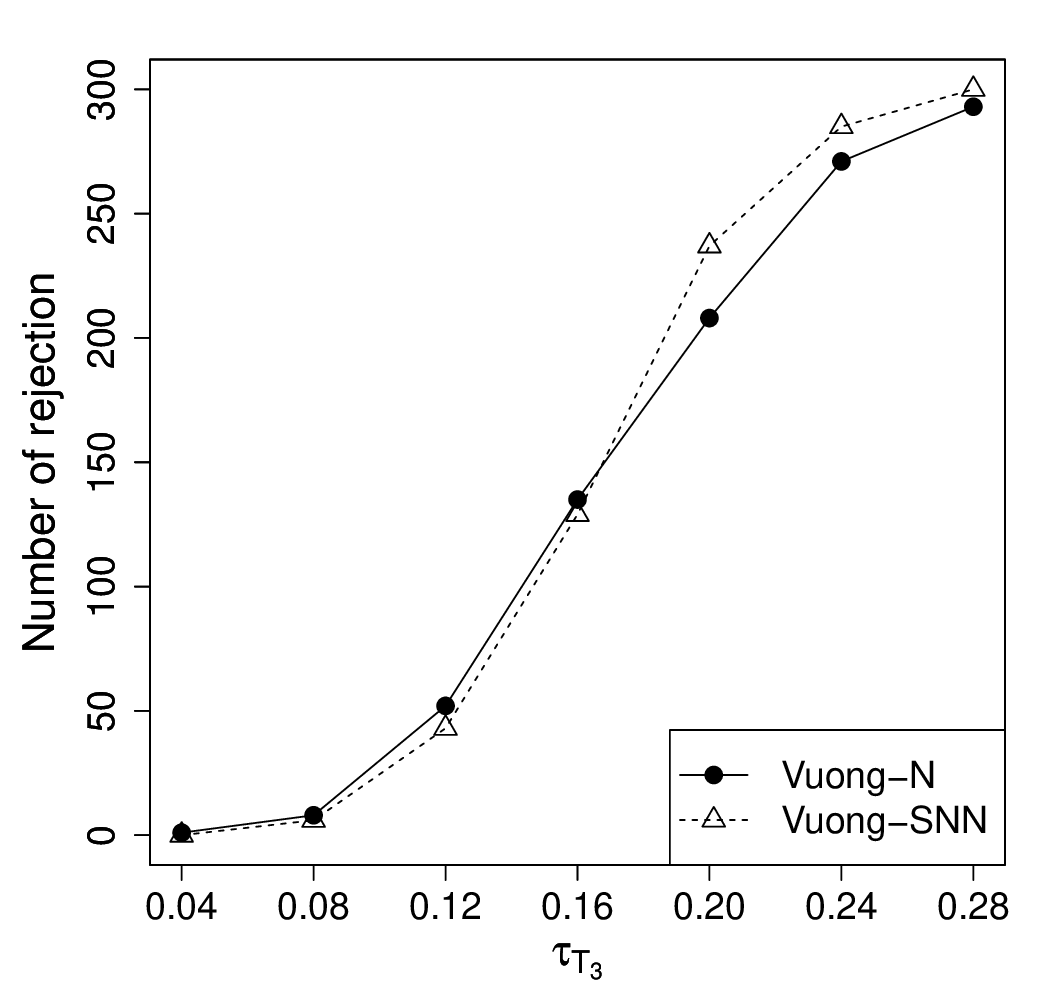}
        \includegraphics[height=6cm]{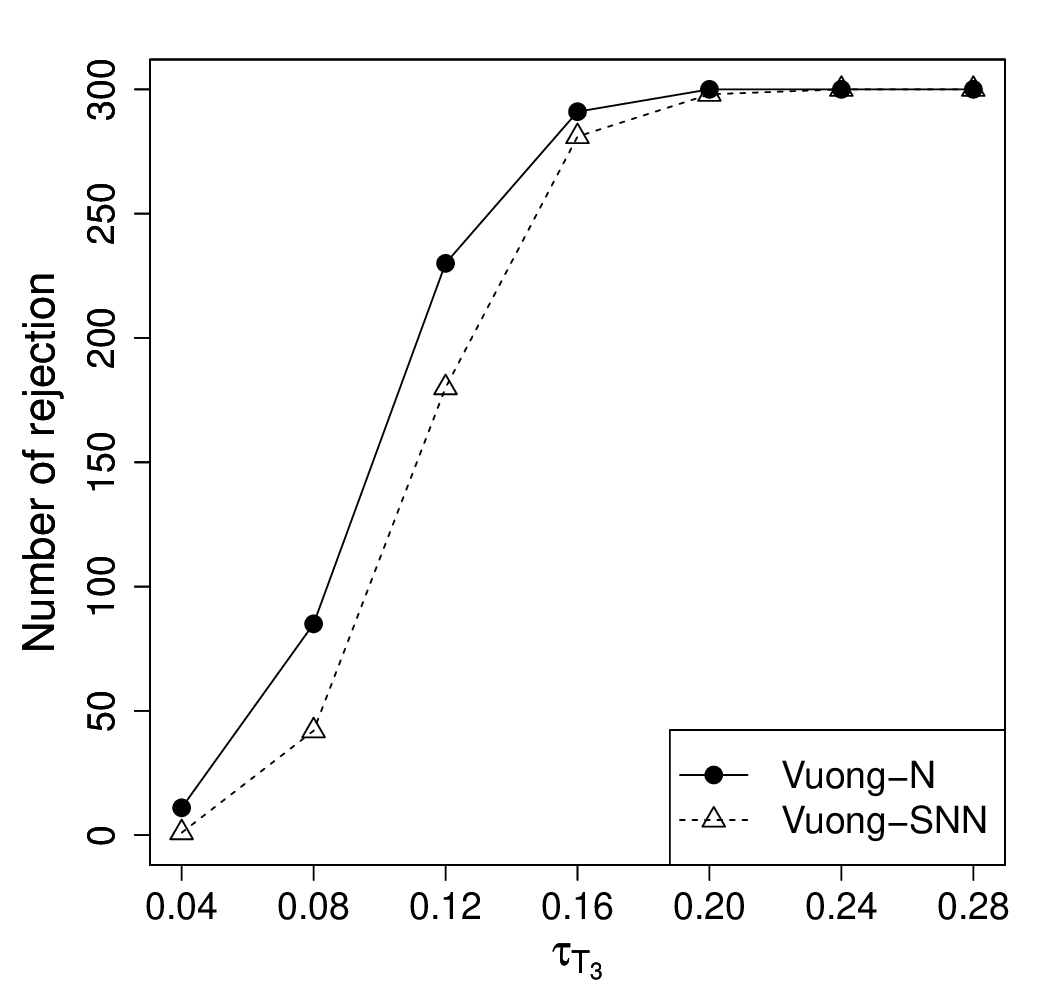}\\
        \includegraphics[height=6cm]{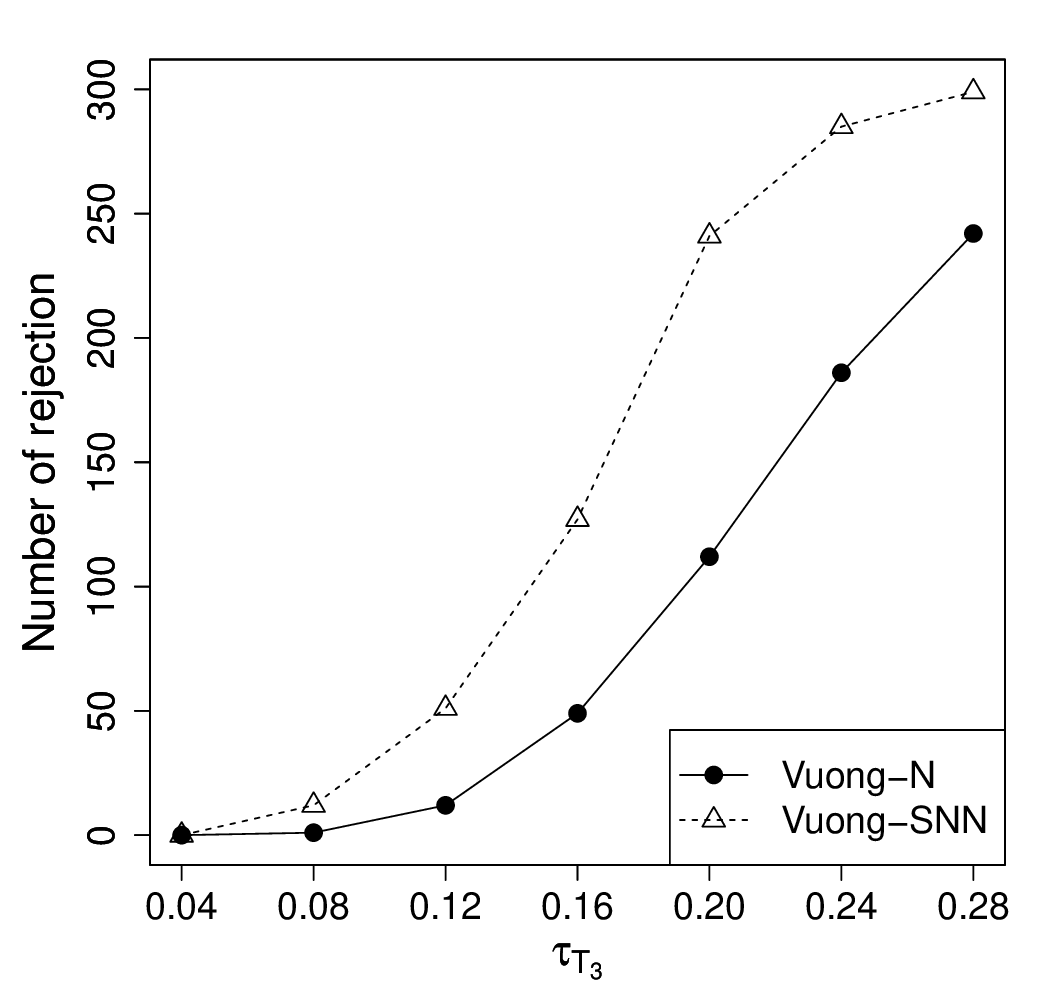}
        \includegraphics[height=6cm]{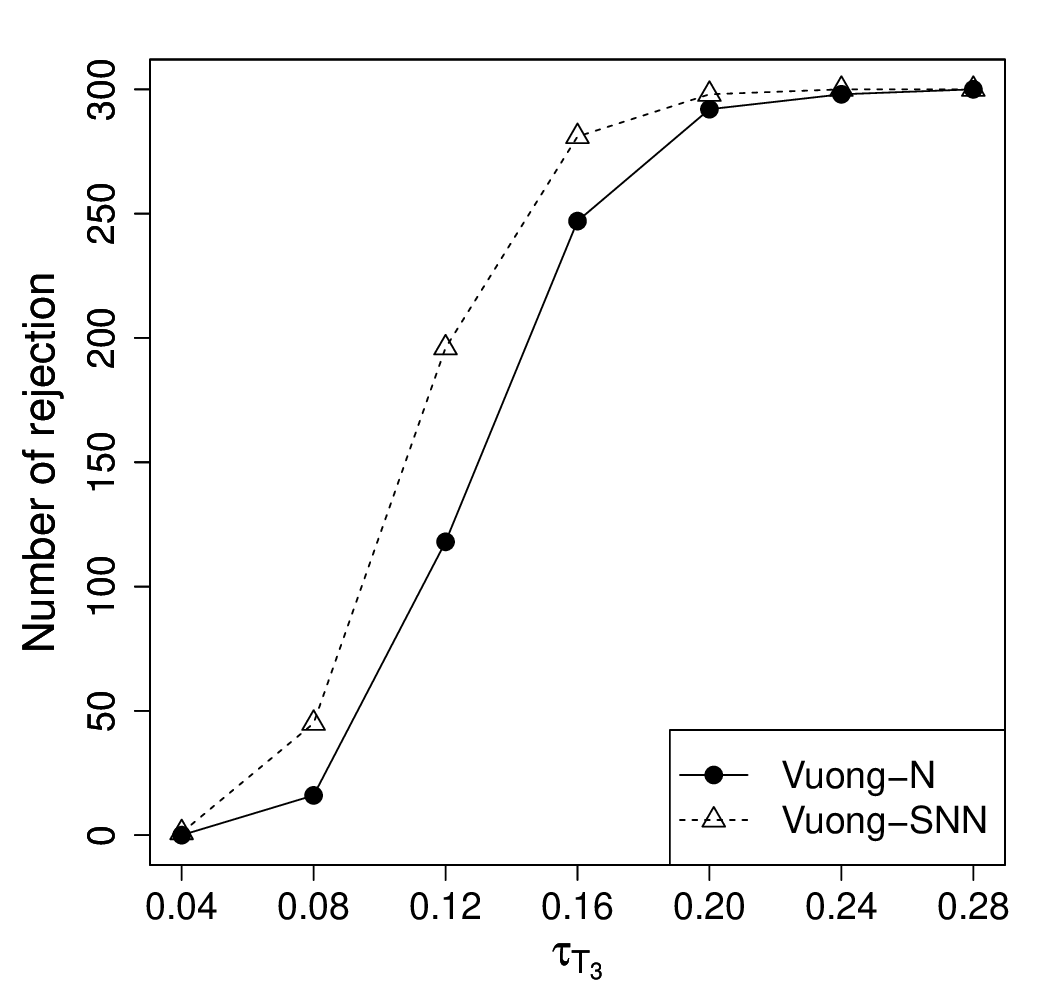}
        \caption{The number of rejections by the Vuong test $(\alpha=0.05)$ comparing \(\bm G^2_{\bm\gamma_2}\) and \(\bm F_{\bm\theta}\) for \(\tau_{T_1} = 0.12\) and \(\tau_{T_2} = 0.08\) (top) and \(\tau_{T_1} = 0.28\) and \(\tau_{T_2} = 0.28\) (bottom) with \(n = 200\) (left) and \(n = 500\) (right), corresponding to the conditions of Figs. \ref{fig:4d_D_tr2_pval1} and \ref{fig:4d_D_tr2_pval2}.}
        \label{fig:4d_D_tr2_num}
    \end{center}
\end{figure}

\begin{figure}[tbh]
    \begin{center}
        \includegraphics[height=6cm]{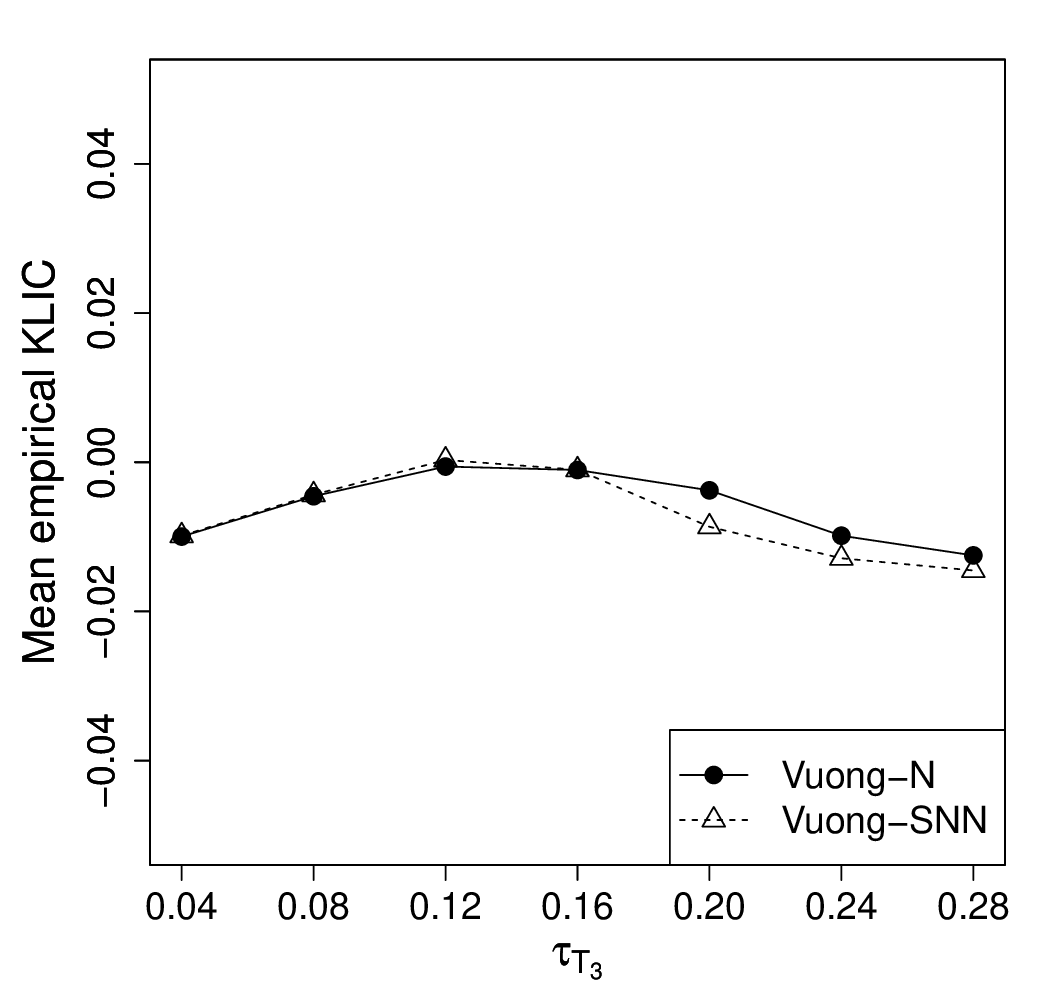}
        \includegraphics[height=6cm]{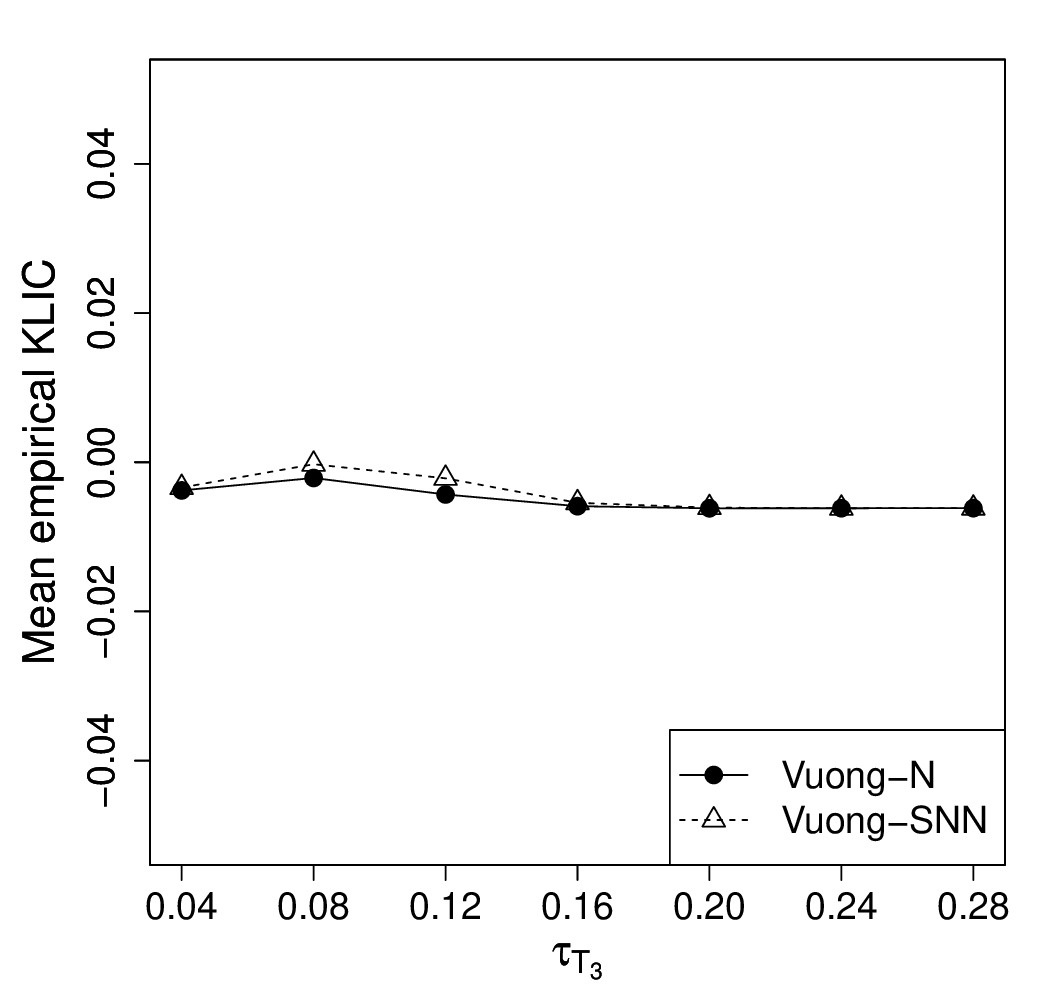}\\
        \includegraphics[height=6cm]{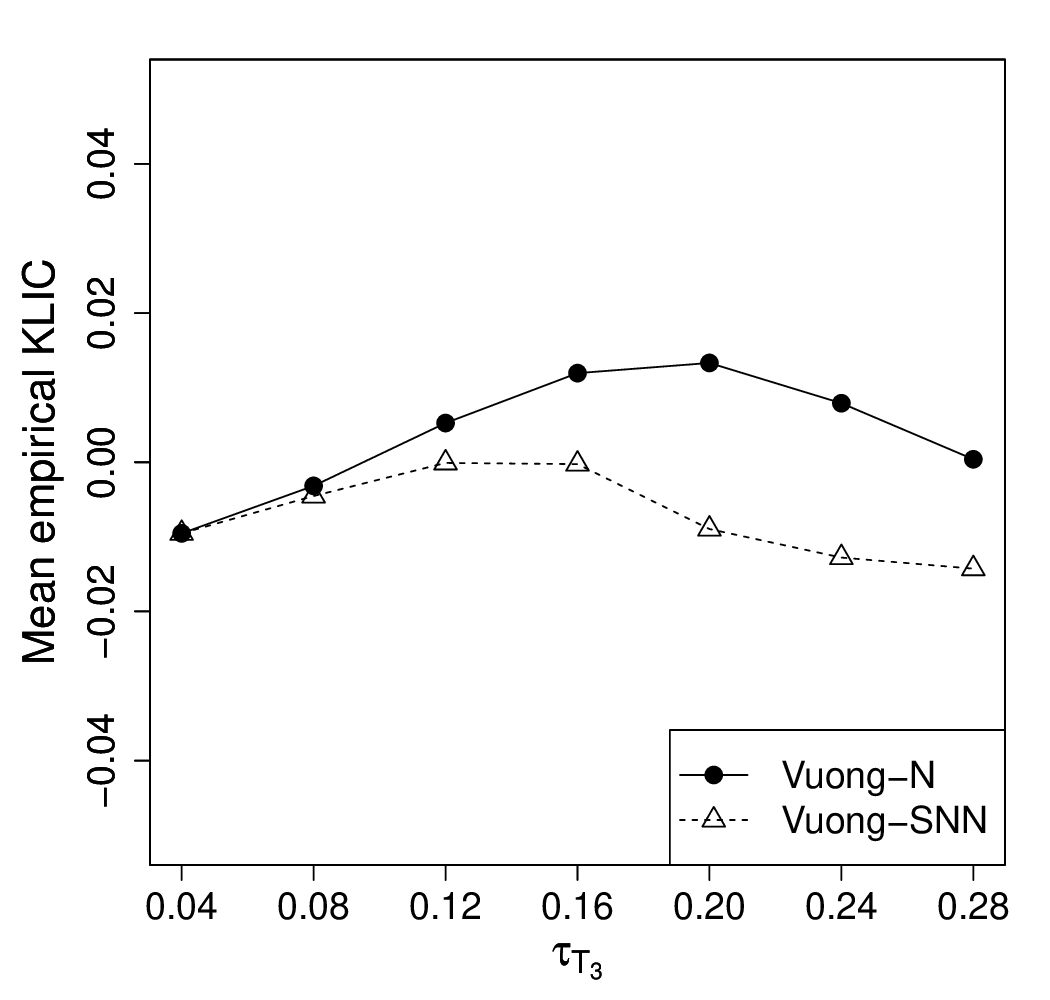}
        \includegraphics[height=6cm]{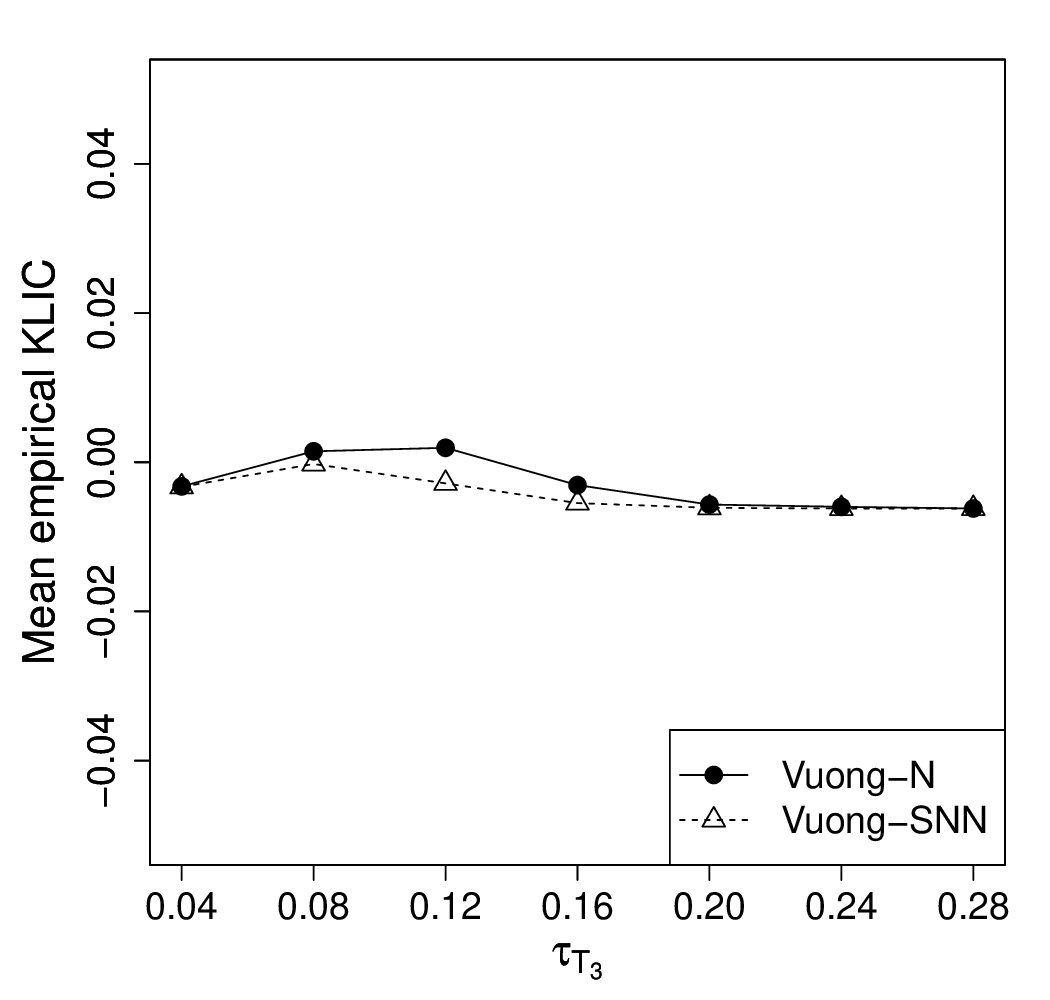}
        \caption{Mean empirical KLIC of the models chosen by Vuong test $(\alpha=0.05)$ comparing \(\bm G^2_{\bm\gamma_2}\) and \(\bm F_{\bm\theta}\) with respect to $h_0$ for \(\tau_{T_1} = 0.12\) and \(\tau_{T_2} = 0.08\) (top) and \(\tau_{T_1} = 0.28\) and \(\tau_{T_2} = 0.28\) (bottom) with \(n = 200\) (left) and \(n = 500\) (right), corresponding to the conditions of Fig. \ref{fig:4d_D_tr2_num}.}
        \label{fig:4d_D_tr2_KLIC}
    \end{center}
\end{figure}

The p-values from the Vuong-N and Vuong-SNN tests comparing \(\bm G^2_{\bm\gamma_2}\) and \(\bm F_{\bm\theta}\) for \(\tau_{T_1} = 0.12\) and \(\tau_{T_2} = 0.08\) are shown in Fig. \ref{fig:4d_D_tr2_pval1}. Under these weaker correlation conditions, the median of p-values obtained by Vuong-N test remained lower than Vuong-SNN test in overall, except for \(\tau_{T_3} \geq 0.20\) with \(n=200\), where the median of p-values obtained by Vuong-N test remained slightly larger than Vuong-SNN test. Conversely, for \(\tau_{T_1} = 0.28\), \(\tau_{T_2} = 0.28\) and \(\tau_{T_3} \geq 0.08\), the median of p-values obtained by Vuong-N test was consistently higher, as shown in Fig. \ref{fig:4d_D_tr2_pval2}. Figures \ref{fig:4d_D_tr2_num} and \ref{fig:4d_D_tr2_KLIC} present the corresponding rejection counts and mean empirical KLIC. Notably, for \(\tau_{T_1} = 0.28\) and \(\tau_{T_2} = 0.28\) with \(n=200\), the Vuong-SNN test showed significantly higher rejection rates near \(\tau_{T_3} \approx 0.20\), while the difference almost diminished when \(n=500\). The mean empirical KLIC suggests superior model selection performance of Vuong-SNN test under \(\tau_{T_1} = 0.28\), \(\tau_{T_2} = 0.28\), and \(\tau_{T_3} \geq 0.12\) with \(n=200\), in alignment with the p-value and rejection count results.


\section{Application}

We analyze historical daily log-returns of three financial instruments observed from 3rd January 2012 to 30th December 2013, all made of $n=511$ observations:  
\begin{itemize}
    \item The British Pound (GBP) to United States Dollar (USD) exchange rate (GBP/USD);
    \item The World Gold Council gold price in USD per troy ounce (GOLD);
    \item The German Stock Index (DAX).
\end{itemize}
The data were obtained from the R package \texttt{qrmdata} \citep{qrmdata}.
Prior to analyzing dependence in the dataset, we filtered the returns using a GARCH(1,1) model with a constant mean assumption. The model's error distribution was specified as a Student's $t$-distribution. The standardized residuals were then transformed into uniform pseudo-observations using their empirical distribution functions.

We fitted the models \(\bm{F}_{\bm{\theta}}\) and \(\bm{G}_{\bm{\gamma}}\) to the pseudo-observations using the R package \texttt{VineCopula} \citep{VineCopula}. Subsequently, we applied the Vuong-N and Vuong-SNN tests to these models using the R package \texttt{nonnest2} \citep{nonnest2}, following a procedure similar to that employed in the simulation studies. The resulting p-values were \(0.039\) for Vuong-N test and \(0.18\) for Vuong-SNN test. At a significance level of \(\alpha = 0.05\), these findings suggest that Vuong-N test distinguishes between the two competing models, while Vuong-SNN test does not.

\section{Conclusion}

This paper has investigated the model selection tests in vine copula truncation settings, focusing on the effectiveness of Vuong test under both nested and strictly non-nested hypotheses, referred to as Vuong-N and Vuong-SNN tests, respectively. Through extensive simulation studies in three- and four-dimensional settings, we evaluated the performance of the Vuong-N and Vuong-SNN tests using p-values, the number of rejections, and mean empirical KLIC. The results reveal that the relative performance of each test is sensitive to the strength of dependencies within the vine structure. In scenarios with weaker pairwise dependencies, the nested hypothesis is more likely to hold, and Vuong-N test produced lower p-values and higher rejection rates, along with improved mean empirical KLIC. Conversely, when the dependencies are stronger, Vuong-SNN test yielded valid and often superior model distinctions, demonstrating that strictly non-nested testing, despite its heuristic status, remains an informative approach in such settings.

These findings indicate that neither test should be considered universally superior. Rather, the appropriateness of nested versus strictly non-nested hypotheses should be assessed in light of the dependency characteristics of the data.
For practitioners, our results underscore the importance of aligning model selection tests with the underlying properties of the data. By characterizing the conditions under which nested hypotheses enhance test power and clarifying the continued utility of the strictly non-nested framework, this study provides a broader perspective on model selection for truncated vine copulas. Our findings offer both methodological insight and practical guidance, laying the groundwork for future refinement of model selection strategies in multi-dimensional dependence modeling.

\subsection*{Computational Environment}

All simulations were performed using R 4.4.2 with the \texttt{doParallel} package for parallel computing, on a Windows 10 machine equipped with an AMD Ryzen Threadripper 3990X 64-Core Processor and 256~GB of RAM. The total computation time was approximately one week.

\subsection*{Funding}

Ichiro Nishi gratefully acknowledges the financial support from the Graduate University for Advanced Studies (SOKENDAI) under SOKENDAI Student Dispatch Program grant. Yoshinori Kawasaki is supported by JSPS Grants-in-Aid for Scientific Research (23K25506).

\section*{Declarations}

\subsection*{Competing interests}

The authors have no competing interests to declare that are relevant to the content of this article.

\bibliographystyle{plainnat}
%\bibliography{main}

\end{document}